\documentclass[traditabstract]{aa}  
\usepackage{graphicx}
  \usepackage[fleqn]{amsmath}
\usepackage{txfonts}
\usepackage{natbib}
\bibpunct{(}{)}{;}{a}{}{,}	
\begin{document}
   \title{Resolving the fragmentation of high line-mass filaments with ALMA: the integral shaped filament in Orion A}
      \author{J. Kainulainen\inst{1}, 
           A. M. Stutz\inst{1, 2}, 
           T. Stanke\inst{3},
           J. Abreu-Vicente\inst{1},
           H. Beuther\inst{1}, 
           T. Henning\inst{1},
           K. G. Johnston\inst{4}, \and
           S. T. Megeath\inst{5}
          }


   \institute{Max-Planck-Institute for Astronomy, K\"onigstuhl 17, 69117 Heidelberg, Germany \\
              \email{jtkainul@mpia.de}     
              \and    
              Departamento de Astronom\'ia, Universidad de Concepci\'on, Av. Esteban Iturra s/n, Distrito Universitario, 160-C, Chile
              \and
		ESO, Karl-Schwarzschild-Strasse 2, 85748 Garching bei M\"unchen, Germany   
		\and
		School of Physics \& Astronomy, E.C. Stoner Building, The University of Leeds, Leeds LS2 9JT, UK 
		\and
		Ritter Astrophsical Research Center, Department of Physics and Astronomy, University of Toledo, Toledo, OH 43606, USA \\
		                    }
   \date{Received ; accepted }
  \abstract
{We study the fragmentation of the nearest high line-mass filament, the integral shaped filament (ISF, line-mass $\sim$ 400 M$_\odot$ pc$^{-1}$) in the Orion A molecular cloud. We have observed a 1.6 pc long section of the ISF with the Atacama Large Millimetre/submillimeter Array (ALMA) at 3 mm continuum emission, at a resolution of $\sim$$3\arcsec$ (1\,200 AU). We identify from the region 43 dense cores with masses about a solar mass. 60\% of the ALMA cores are protostellar and 40\% are starless. The nearest neighbour separations of the cores do not show a preferred fragmentation scale; the frequency of short separations increases down to 1\,200 AU. We apply a two-point correlation analysis on the dense core separations and show that the ALMA cores are significantly grouped at separations below $\sim$17\,000 AU and strongly grouped below $\sim$6\,000 AU. The protostellar and starless cores are grouped differently: only the starless cores group strongly below $\sim$6\,000 AU.  In addition, the spatial distribution of the cores indicates periodic grouping of the cores into groups of $\sim$30\,000 AU in size, separated by $\sim$50\,000 AU. The groups coincide with dust column density peaks detected by \emph{Herschel}. These results show hierarchical, two-mode fragmentation in which the maternal filament periodically fragments into groups of dense cores. Critically, our results indicate that the fragmentation models for lower line-mass filaments ($\sim$ 16 M$_\odot$ pc$^{-1}$) fail to capture the observed properties of the ISF. We also find that the protostars identified with \emph{Spitzer} and \emph{Herschel} in the ISF are grouped at separations below $\sim$17\,000 AU. In contrast, young stars with disks do not show significant grouping. This suggests that the grouping of dense cores is partially retained over the protostar lifetime, but not over the lifetime of stars with disks. This is in agreement with a scenario where protostars are ejected from the maternal filament by the slingshot mechanism, a model recently proposed for the ISF by Stutz \& Gould. The separation
distributions of the dense cores and protostars may also provide an evolutionary tracer of filament fragmentation.} 
  \titlerunning{Resolving the fragmentation of high line-mass filaments} 
  \authorrunning{J. Kainulainen et al.} 
  \maketitle


\section{Introduction} 
\label{sec:intro}


Filamentary structures are fundamental building blocks of the molecular clouds of the interstellar medium (ISM), manifesting themselves over wide ranges of sizes ($\sim$0.1-100 pc), masses ($\sim$1-$10^5$ M$_\odot$), and line-masses ($\lesssim$1\,000 M$_\odot$ pc$^{-1}$) \citep[e.g.,][]{bal87, hac13, alv14, kai13, kai16, abr16}. Specifically, filaments that have line-masses greatly in excess to the critical value of the self-gravitating, thermally supported, non-magnetised, infinitely long equilibrium model, i.e., $\gg$16 M$_\odot$ pc$^{-1}$ \citep[][]{ost64}, contain large enough mass reservoirs to give birth to high-mass stars and star clusters \citep[e.g.,][]{pil06, beu10, beu15a, hen10, sch12, kai13, stu15prep, con16}. This makes understanding their fragmentation and gravitational collapse important for Galactic-scale star formation. We refer to these filaments as \emph{highly super-critical filaments} in this paper. The physics governing the evolution of the highly super-critical filaments may be radically different from those of near-critical filaments, especially due to their strong global gravitational potential that may crucially affect their evolution \citep{stu15prep}. Therefore, the fragmentation properties of highly super-critical filaments should not be extrapolated from studies of near-critical filaments, but dedicated studies are necessary.

The main obstacle in studying the fragmentation of highly super-critical filaments is the observational challenge. They show fragmentation down to (at least) $\sim$1\,500 AU scales \citep[e.g.,][]{tak13, tei15} and can be several parsecs long. This means that building a complete view of their structure requires high resolution mapping over a large area. Coupling the gas structure with star formation also requires an accurate census of the young {(proto-) stellar} population of the cloud; this limits the possible targets to distances closer than $\sim$1 kpc where such a census can be attained \citep[e.g.,][]{eva09, gut11, meg12, stu13, meg16}. 


Currently, the above observational challenges limit the number of possible targets to exactly one: the "integral-shaped filament" \citep[ISF; e.g.,][]{bal87, pet08} in the \object{Orion A molecular cloud}. At the distance of 420 pc \citep[e.g.,][]{sch14}, the ISF is the nearest highly super-critical filament, and especially the nearest such filament in a giant molecular cloud that exhibits high-mass star formation. The ISF, located in the northern part of Orion A, is an $\sim$8 pc long structure with the line-mass of $385 \times (d/\mathrm{pc})^{3/8}$ M$_\odot$ pc$^{-1}$ from 0.05 pc < $d$ < 8.5 pc of projected separation from the filament ridge \citep{stu15prep}. Importantly, the young stellar population of Orion A is well characterised with \emph{Spitzer} \citep{meg12} and \emph{Herschel} \citep[the \emph{Herschel} Orion Protostar Survey, HOPS;][]{fur16}. Combined, these properties make the ISF an outstanding, and currently unique, region in which to study the physics of how highly super-critical filamentary gas fragments into (high- and low-mass) protostars. While several other highly super-critical filaments are known \citep[e.g.,][]{jac10, hil11, kai13, abr16, hen16}, their larger distances do not allow an analysis as detailed as can be performed in the ISF. 

While the Orion A cloud has been a target of a myriad of studies \citep[see Fig. 1 in][]{mei16}, the sensitivity and resolution of the interferometer arrays have only recently enabled detailed fragmentation studies of the ISF. \citet{tei15} and \citet{tak13} have analysed fragmentation in sub-parsec-sized sections of the ISF in $\sim$1\,600 AU resolution \citep[for studies in lower spatial resolution, see, e.g.,][]{bal87, joh99, chi97, stu15prep}. These observations have established evidence of quasi-periodic fragment separations and hierarchical fragmentation in the filament. However, the relatively small areas covered by the surveys have hampered the statistical analyses of their fragment distributions. Similarly, no work so far has directly linked the distribution statistics of fragments to those of the young stellar population in the ISF.

We report in this paper the most sensitive fragmentation study of a highly super-critical filament to date, based on our ALMA interferometer study of the ISF and the census of the young stellar population derived by HOPS. Our mapping more than  doubles the area of the ISF studied previously in a similar resolution, also reaching higher sensitivity. Our specific goals in this paper are to determine the fragmentation length scales of the filament, to look for preferential fragmentation scales and grouping of fragments, and to establish the relationship of the fragmentation scales to the distribution of the young stellar population in the ISF. 

\section{Observations and methods} 
\label{sec:data}

\begin{figure*}
   \centering
\includegraphics[width=\textwidth]{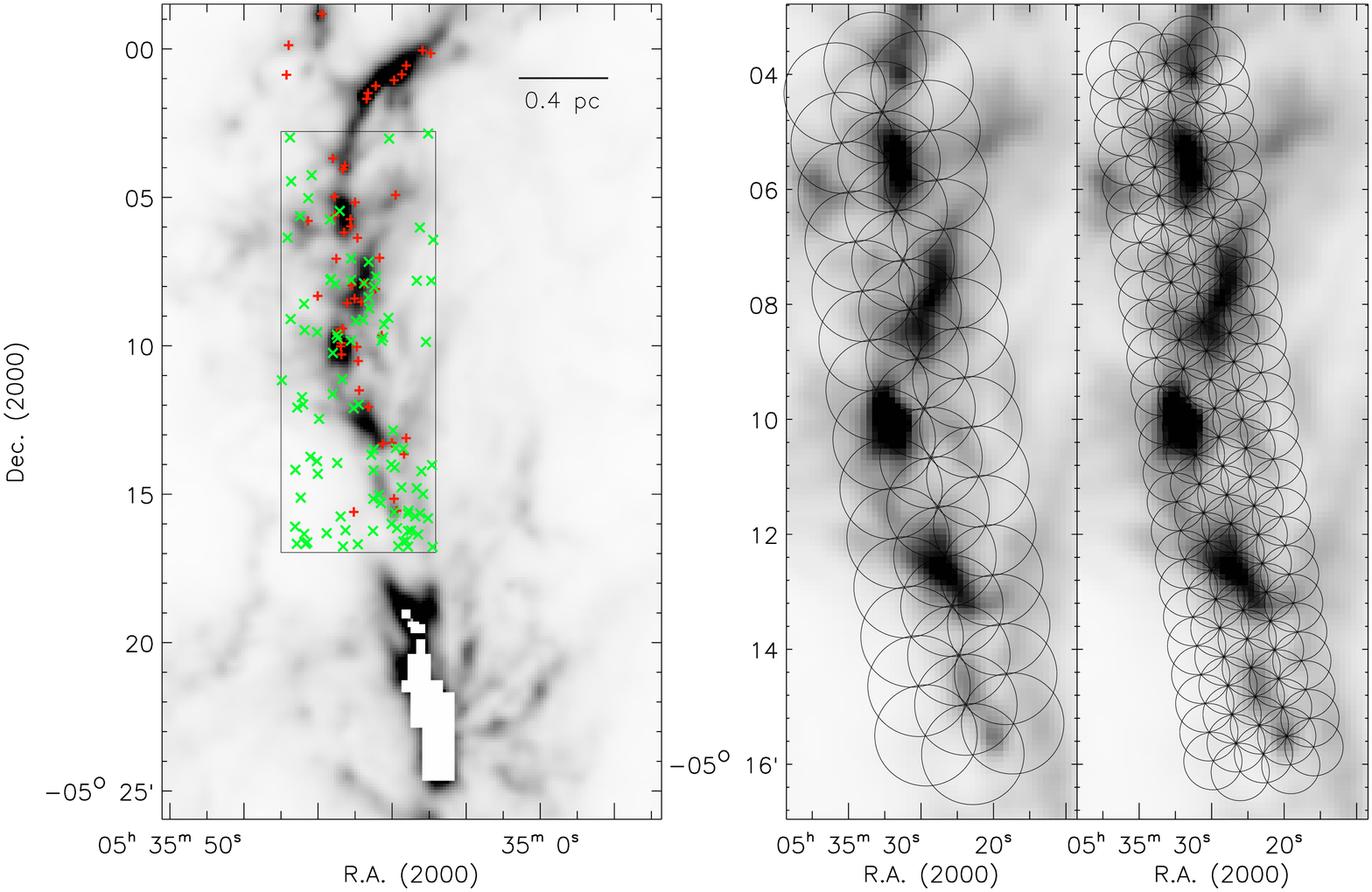}
      \caption{\emph{Left: }column density map of the northern part of the ISF, derived from \emph{Herschel} data \citep{stu15}. The rectangle indicates the location of the zoom-ins shown in the right panel. The red plusses show the locations of protostars and green crosses the locations of stars with disks \citep{meg12, fur16}. For visibility, only disks within the squared area are shown. The white pixels result from saturation of the detector. \emph{Right: }Zoom-in of the left panel showing the 7-m array and 12-m array mosaic footprints. 
              }
         \label{fig:footprint}
   \end{figure*}

We used the ALMA interferometer during Cycle 2 to observe the northern part of the ISF at 3 mm continuum emission and C$^{18}$O line emission (ESO Project ID 2013.1.01114.S, PI Kainulainen). In this paper we only use the continuum emission data. Both the 12-m and the 7-m (Atacama Compact Array; ACA) arrays were employed. The 7-m array observations were conducted on 16th August 2014 and the 12-m array observations in a compact configuration on 16th January 2015. 
We covered a region $\sim$$3 \times 11\arcmin$ ($\sim$$0.37 \times 1.34$ pc) in size in \object{OMC-2} and \object{OMC-3} with mosaicing observations. The footprints of the 12-m and 7-m arrays are shown in Fig. \ref{fig:footprint}. During the observations, three of the four correlator sidebands, each 2 GHz wide, were set to central frequencies of 107.29217 GHz, 96.97848 GHz, and 98.97848 GHz. The fourth sideband was set to the frequency of the C$^{18}$O line at 109.749042 GHz. The relevant observational parameters are listed in Table \ref{tab:observations}.


\begin{table}
\caption{Parameters of the ALMA observations}             
\label{tab:observations}     
\centering                    
\begin{tabular}{l c c}     
\hline\hline            
		Parameter		& 12-m array    			& 7-m array  	\\
\hline
Primary beam			& $\sim$60$\arcsec$		& $\sim$105$\arcsec$	\\
Max recoverable scale\tablefootmark{a}  &                   25\arcsec                          &	 42\arcsec	\\
Synthesized beam		& $3\farcs2 \times 1\farcs7$			&	$19\farcs 9 \times 10\farcs 3$ \\
Beam position angle		& 	71$^\circ$			& 76$^\circ$  \\
No. of mosaic points   	& 127   				&  42			\\
On-source time 		& 78 min				& 3h 44 min	\\      
Baseline length range	& 15.1-348.5 m			& 8-48.9 m 	\\
					&  5.0-116.2 k$\lambda$		&  2.7-16.3 k$\lambda$				\\
Achieved $rms$ 		& 0.14 mJy beam$^{-1}$	& 2.3 mJy beam$^{-1}$	\\
\hline
\end{tabular}
\tablefoot{
\tablefoottext{a}{According to the ALMA Cycle 2 Proposer's Guide.}
}
\end{table}


During the 12-m array observations, J0423-0120 and Uranus were observed in the beginning of the measurement set for bandbass and flux calibration, respectively. J0517-0520 was observed every $\sim$10 min during the science observations for phase calibration. A similar sequence was employed during the 7-m array observations, using J0607-0834 for bandbass calibration, Uranus or J0423-013 for flux and amplitude calibration, and J0541-0541 for phase calibration. 

Both the 7-m and 12-m array data were calibrated and imaged using the \textsf{casa} version 4.5.0. The 12-m array data were calibrated using the ALMA calibration pipeline (part of \textsf{casa} 4.5.0). The 7-m data were calibrated using the calibration scripts prepared by the ALMA Regional Center experts; these are included in the standard data delivery and they available in the archive with the data. 


We imaged the 7-m and 12-m array data both separately and simultaneously using the \textsf{clean} task of \textsf{casa}. Prior to cleaning, the relative weights of the 7-m and 12-m array visibility data were adjusted with the \textsf{casa} task \textsf{statwt}. We excluded from the cleaning the channels at the edges of the bands and the channels coinciding with prominent emission lines. The total bandwidth used in the imaging was 11.1 GHz (including two polarizations). 

During cleaning, we used the `natural' weighting scheme. We first ran the \textsf{clean} task without any masking of emission; the resulting image of this first cleaning was used to make the cleaning mask; this was done by setting boxes around significant emission structures in the cleaned map. The cleaning was then repeated with this mask, yielding the final, cleaned map. This procedure was repeated for 7-m array and 12-m array separately, and also for the two arrays together. The final maps are shown in Fig. \ref{fig:results}. The final synthesised beam size full width half max, $FWHM$) of the combined 7-m and 12-m array map is $3\farcs75 \times 2\farcs27$ ($\sim$1\,600 AU $\times$ 950 AU) and the beam position angle 71$^\circ$. The root-mean-square ($rms$) noise measured from an approximately emission free area is $\sigma_\mathrm{rms}$=0.23 mJy beam$^{-1}$. We use this combined map in the remainder of this paper and refer to it as the ALMA map.

\begin{figure}
   \centering
\includegraphics[width=\columnwidth]{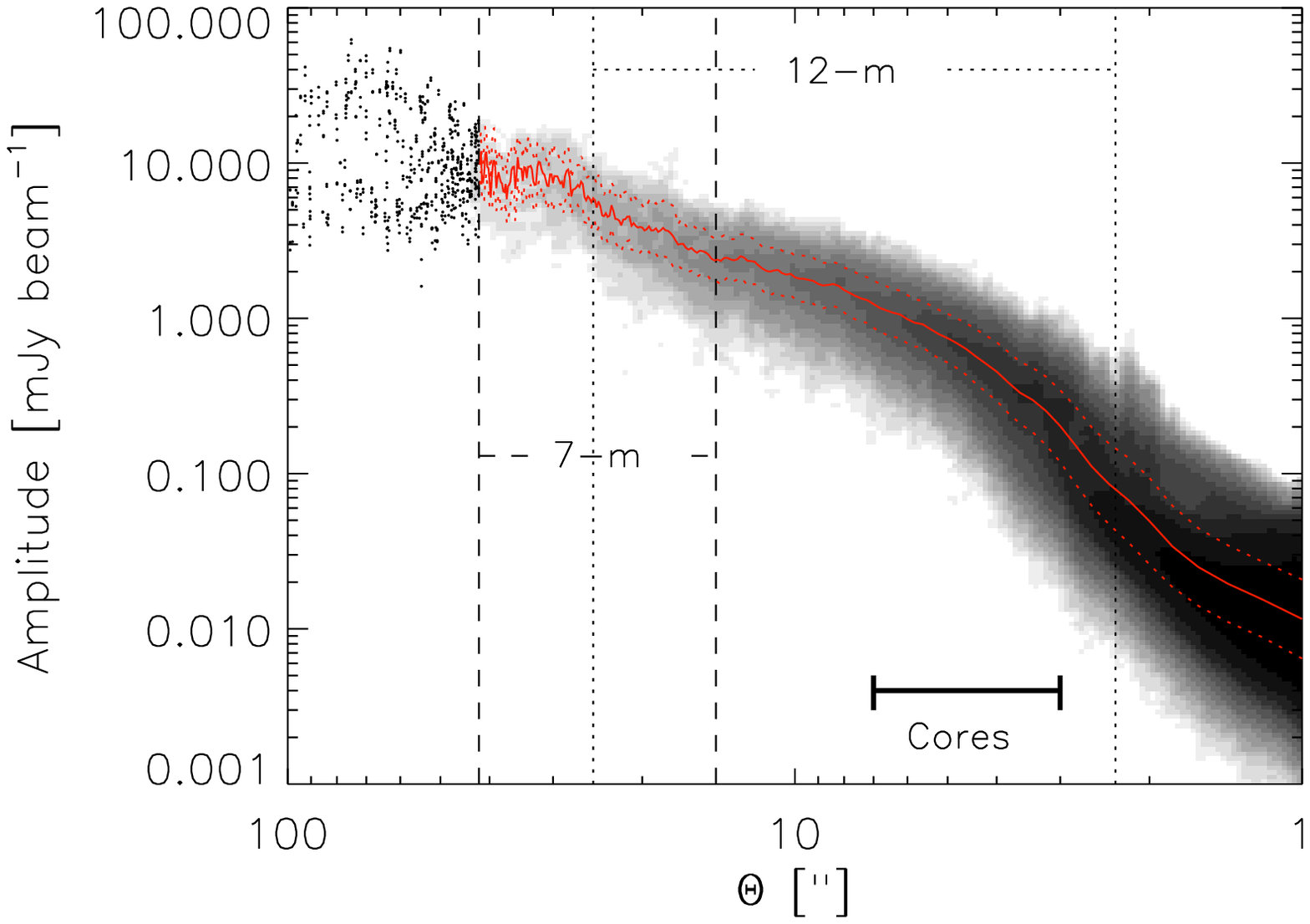}
\includegraphics[width=\columnwidth]{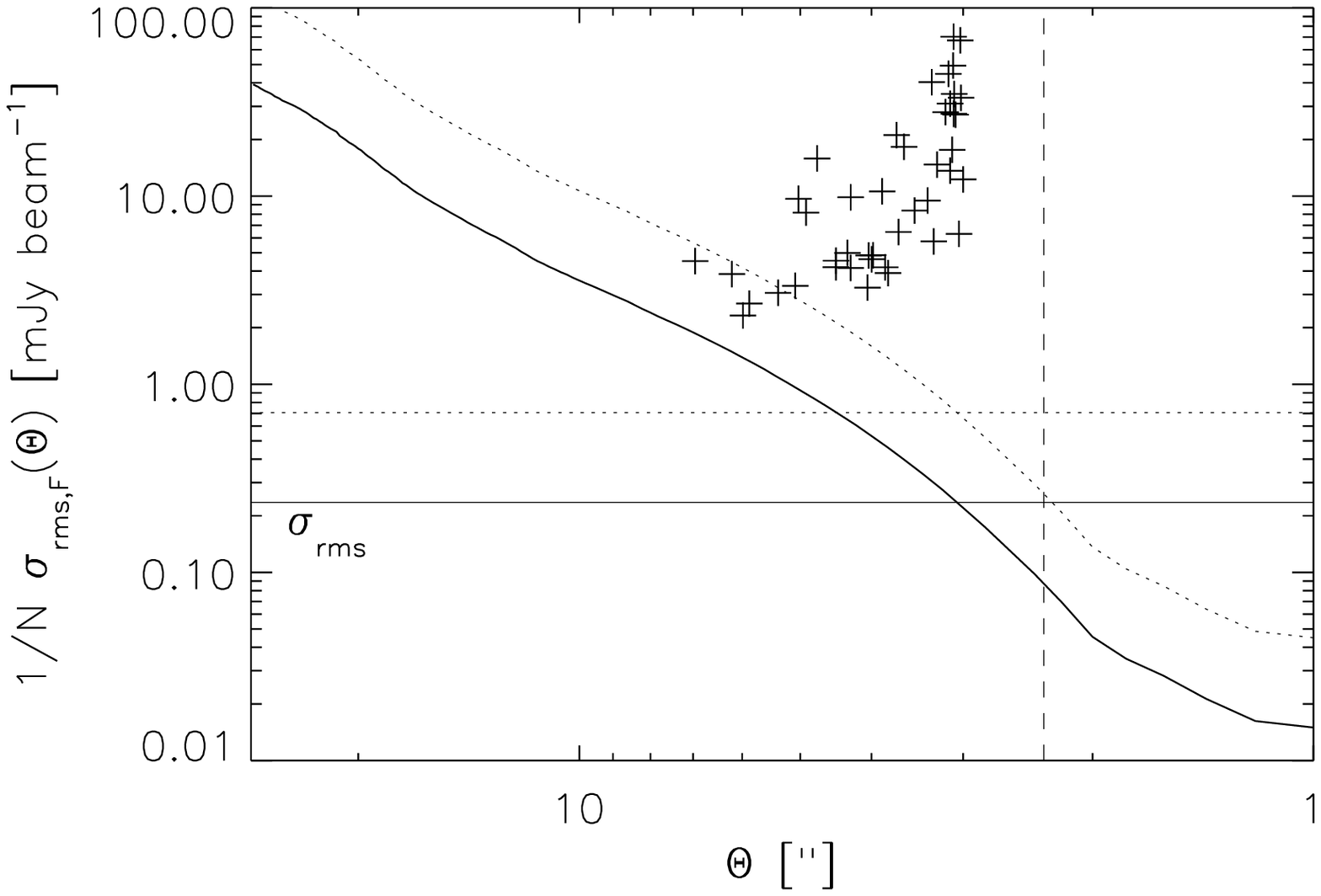}
      \caption{\emph{Top: }Amplitudes of the Fourier modes of the ALMA map. The spatial scales probed by the 12-m and 7-m arrays are indicated with vertical lines. The size range of the identified cores is shown with a horizontal line. The mean spectrum is shown with a solid red line and the standard deviation with dotted red lines. For visibility, only data points above 25\arcsec are shown; at smaller scales the grey scale shows the density of data points. \emph{Bottom: }The $rms$ noise measured in the Fourier domain, i.e., $1/N \ \sigma_\mathrm{rms, F}(\Theta)$. The dotted line shows the $3\sigma_\mathrm{rms, F}(\Theta)$ value. The horizontal lines show one (solid line) and three times (dotted line) the standard deviation measured from an apparently empty region of the ALMA map, i.e., $\sigma_\mathrm{rms}$. The dashed vertical line indicates the beam size. The plusses indicate the 43 compact objects identified from the ALMA map; the y-position is determined by the peak flux and the x-position by the geometric radius of the objects.
              }
         \label{fig:powerspectrum}
   \end{figure}
\begin{figure*}
   \centering
\includegraphics[bb = 100 10 745 680, clip=true, width=\textwidth]{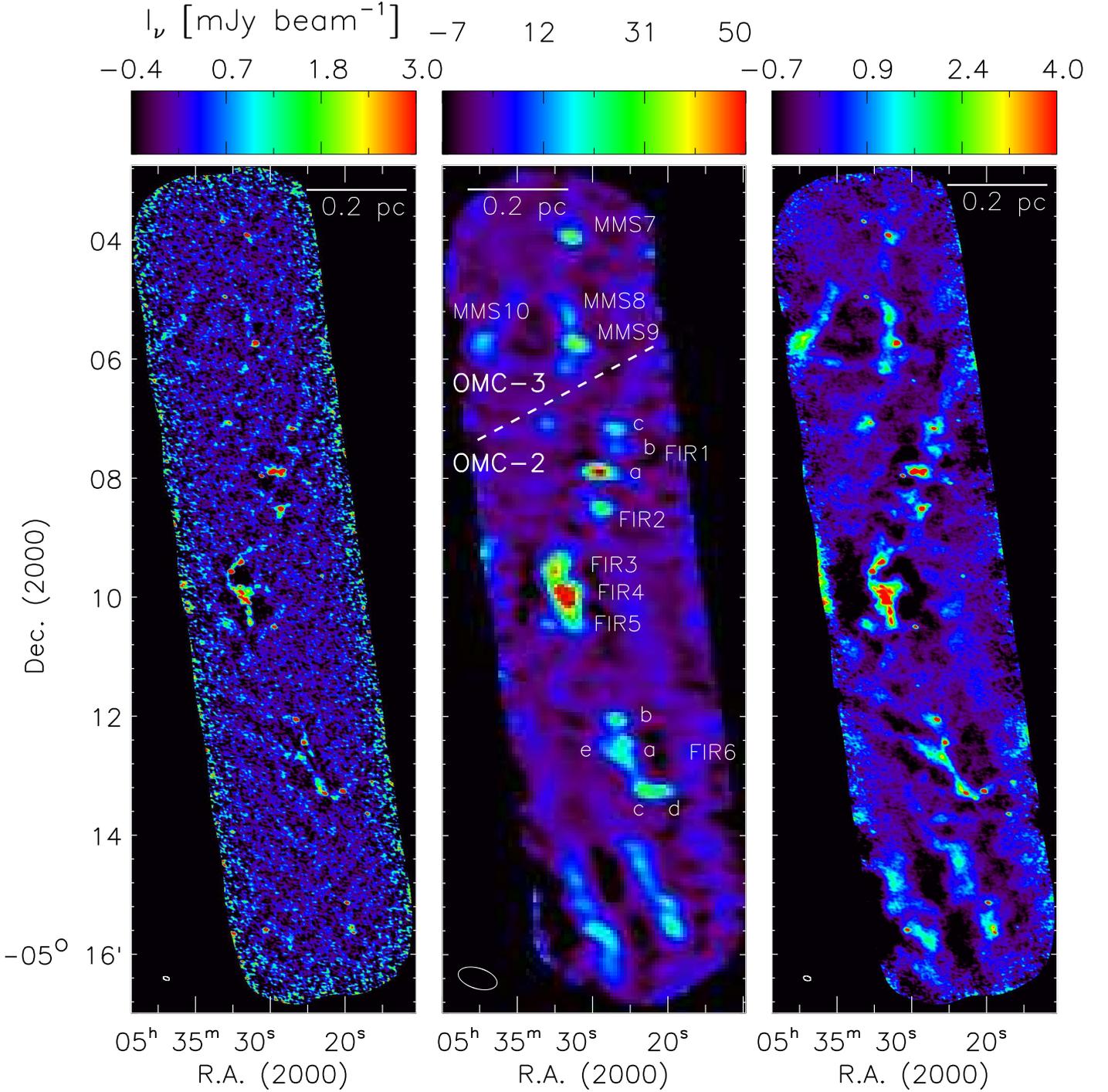}
      \caption{3 mm continuum emission maps of the OMC-2/3 region, observed with ALMA. The $FWHM$ beam is shown in the lower left corner of the frames. \emph{Left: }The map observed with the 12-m array. The beamsize is $3\farcs 2 \times 1\farcs7$. \emph{Center: }The map observed with the 7-m array. The beamsize is $19\farcs 9 \times 10\farcs 3$. The region designations from \citet{mez90} and \citet{chi97} are shown in the frame. \emph{Right: }The map combining the 7-m and 12-m array data. The beamsize is $3\farcs75 \times 2\farcs27$. See Fig. \ref{fig:results3} for higher resolution zoom-ins of the map.
              }
         \label{fig:results}
   \end{figure*}

We further investigate the sensitivity of the ALMA map by inspecting the amplitudes of its Fourier modes. Consider the ALMA map, $f(x,y)$, its dimensions, $S_\mathrm{x}$ and $S_\mathrm{y}$, and its discrete Fourier transform, $F(n, m)$. The transformation between the angular scale, $\Theta$ [$\arcsec$], and the Fourier modes is then
\begin{equation}
\Theta  = \bigg(  (\frac{n}{S_\mathrm{x}})^2  + (\frac{m}{S_\mathrm{y}})^2   \bigg)^{-1/2}. 
\end{equation}
Figure \ref{fig:powerspectrum} shows the amplitudes of the Fourier modes as a function of the angular scale. The figure also shows the mean and standard deviation of the amplitudes, calculated within a $0.5\arcsec$ wide window. To improve visibility, only the amplitudes at scales larger than $42\arcsec$ (the filtering scale of the 7-m array) are plotted; at smaller scales the density of points is shown instead. The figure demonstrates the quality of the data: between $\sim 5-30 \arcsec$, the relationship is smooth and power-law-like. Below about $5\arcsec$ the relationship steepens, evidently due to beam-size effects. Above 42$\arcsec$, the relationship is not well-behaved, which indicates severe filtering effects. These effects manifest themselves in the image plane, e.g., as negative bowls (e.g., surrounding the FIR3-5 complex). 

We also used the Fourier amplitudes to empirically estimate the $rms$ noise of the ALMA map as a function of the spatial scale. We measured the $rms$ of the Fourier spectrum as a function of the spatial scale, $\sigma_\mathrm{rms, F}(\Theta)$, within a window that has the width of $\Theta$. Note that the choice of the window size is somewhat arbitrary; the resulting $rms$ should be considered a rough estimate. Then, an estimate of the image $rms$ is $1/N \times \sigma_\mathrm{rms, F}(\Theta)$, where $N$ is the number of data points, i.e., of Fourier modes (results from Parseval's theorem). The scale-dependent $rms$, $1/N \times \sigma_\mathrm{rms, F}(\Theta)$, is shown in Fig. \ref{fig:powerspectrum}, lower panel. We discuss this relationship further in the context of structure identification later in the paper.

We also employ in this work the column density map of the Orion A molecular cloud derived from \emph{Herschel} far-infrared emission observations by \citet{stu15}. The section of the map covering the ISF is shown in Fig. \ref{fig:footprint}. The map has the spatial resolution of $FWHM=18\arcsec$ (7\,600 AU). We refer to \citet{stu15} for the details of the map.


Finally, we also use the catalog of \emph{Spitzer}-identified protostars and stars with disks from the HOPS survey (\citealt{meg12} for the disks and \citealt{fur16} for the protostars). The protostars and stars with disks are shown in Fig. \ref{fig:footprint}.


\subsection{Identification of dense cores}
\label{subsec:core-detection}


To study the spatial distribution of dense cores in the ALMA map, we first need to identify such objects. Generally, identifying ``cores'' from column density maps is an ill-defined problem, because there is no common standard for what exactly constitutes ``a core'' \citep[see discussion in, e.g.,][]{dif07, berg07, and14}. Consequently, different approaches undoubtedly result in somewhat different results \citep[e.g.,][]{smi08, kai09, bea13}. Here, our emphasis is in taking the advantage of the high resolution of the ALMA data, and hence, in identifying structures that are centrally concentrated at scales a few times the beamsize (i.e., $\lesssim$3\,000-4\,000 AU). The ALMA data recover structures also at larger spatial scales, up to $\sim$20\,000 AU, due to the inclusion of the ACA array (see Fig. \ref{fig:powerspectrum}). Since the ISF shows significant extended structures (e.g., Fig \ref{fig:footprint}), this results in a varying background for the compact structures we wish to identify. Also, some regions of the map clearly contain high surface densities of local maxima. These properties are suited to a technique based on 2-dimensional model fitting (as opposed to segmentation). Finally, we also wish to adopt a well-documented and publicly available algorithm that will be straightforward to use in future studies. 


With the above considerations, we chose to use the implementation of \textsf{gaussclumps} in the \textsf{starlink/cupid} package version 2.0\footnote{Available at \url{http://starlink.eao.hawaii.edu/starlink/CUPID} \citep{ber07}. The \textsf{gaussclumps} was originally developed by \citet{stu90}; see the \textsf{cupid} manual for the minute differences between the \textsf{cupid} implementation and the original version.}. \textsf{gaussclumps} identifies cores by fitting 2-dimensional Gaussians at the locations defined by the highest pixel values in the map. Prior to fitting, a background is subtracted from the map using a median filter. After the peak identification and the fitting of the Gaussian profile, the profile is subtracted from the map and the residual map is used to search for the next object. The most important parameters of the algorithm are the threshold (GaussClumps.Thresh in \textsf{cupid}) defining the minimum acceptable value for the peak of the fitted Gaussian function and the width of the filtering function that is used to subtract the background. We chose to use the threshold three times the $rms$ noise calculated from an emission free region in the ALMA+ACA map, i.e., 0.23 mJy beam$^{-1}$. This value is about two times the $rms$ noise measured empirically from the Fourier spectrum at the size of the beam (Fig. \ref{fig:powerspectrum}). Note that not all structures above the threshold are considered as cores; if the structure profile is not well fitted by a Gaussian, the core is rejected. The filter width for the background subtraction was chosen to be 10$\arcsec$, keeping in mind that we here want to concentrate on detection of objects at scales smaller than a few times the beamsize ($\lesssim$4\,000 AU). This restricts the reasonable choices of the background to roughly 9$\arcsec$-15$\arcsec$ (3-5 times the beamsize). We experimented with different filter widths and found that between 7.5$\arcsec$-15$\arcsec$ the choice of the width does not change the conclusions based on the two-point correlation function (see Appendix \ref{app:bgtest}). 


Running \textsf{gaussclumps} resulted in detection of 43 compact objects from the ALMA map. Figure \ref{fig:powerspectrum} (bottom panel) shows the peak fluxes and mean sizes of the objects. The weakest objects have peak fluxes about 10$\sigma_\mathrm{rms}$, indicating robust detections. Most cores are also above the 3$\sigma_\mathrm{rms, F}(\Theta)$ level; from the five cores below 3$\sigma_\mathrm{rms, F}(\Theta)$ (\#23, 28, 29, 33, and 35), all except \#23 are local emission maxima. The 43 identified objects are listed in Table \ref{table:cores} together with their intensities, integrated fluxes, and sizes (shown in Fig. \ref{fig:results3}). 


We cross-correlated the positions of the cores with the protostars \citep{fur16} and stars with disks \citep{meg12}. The objects were considered coincident if their centroids were within a $2\arcsec$ radius of the protostars or disks. 26 of the objects coincide with a protostar and four objects with a disk source (\#24 that has the identifier HOPS64, and objects \#8, \#34, and \#42, which have no HOPS identifier). These four objects all show extended emission surrounding them. It is possible that they indeed are disks, aligning by chance with emission structures along the line-of-sight. However, it is also possible that they are misclassified protostars. We performed a simple estimation of the probability of chance alignment. Consider the probability of chance alignments of $N_\mathrm{object}$ circular objects and $N_\mathrm{disk}$ point-like disk sources. The probability of having $N$ superpositions is 
\begin{equation}
P(N) = \frac{N_\mathrm{disks}!}{N ! (N_\mathrm{disks}-N)!} p^N (1-p)^{N_\mathrm{disks}-N}, 
\label{eq:probability}
\end{equation}
where $p$ is the probability of one superposition, i.e., $N_\mathrm{cores} \pi R_\mathrm{object}^2 / A$, where $R_\mathrm{object}$ is the radius of the object and $A$ is the total considered area. Using the values $N_\mathrm{cores}$=43, $N_\mathrm{disks}$=51, $R_\mathrm{object}$=2$\arcsec$, and $A$ = $1.5\times10^{5}$ $\square\arcsec$ results in $P$(0)=83\% and the expected value of $N$ of 0.2. Against this expectation, we reclassify the four disks that coincide with ALMA-identified compact objects as protostars and consider them as likely misclassifications in \citet{meg12}. In particular, HOPS 64 (\#24) shows a rising SED between 1.6 $\mu$m and 24 $\mu$m and has been shown to have an outflow cavity in HST 1.6 $\mu$m imaging (J. Booker, priv. communication). The other sources have flat to Class II SEDs between 1.6-24 $\mu$m; these may be protostars observed near a face-on inclination and/or with lower mass envelope \citep{fur16}.

We refer to the 43 ALMA-identified compact objects as dense cores in the remainder of this paper. 17 of the dense cores do not coincide with protostars and are considered to be starless cores. Thus, our dense core catalog consists of 26 protostellar cores and 17 starless cores. We note that we do not assess whether the starless cores are gravitationally bound or not (gravitationally bound cores that do not coincide with protostars are commonly referred to as ``prestellar cores''), but rather consider all starless cores together.


We note two caveats arising from the structure identification technique. Because of the background filtering of \textsf{gaussclumps}, we may be insensitive to low-contrast cores that have sizes of 10$\arcsec$ (4\,000 AU) or larger. Starless cores typically have low density contrasts \citep[e.g.,][]{alv01, kan05, kai07}. This may affect the distribution statistics of starless cores analysed later in the paper. Protostellar cores are centrally concentrated and are unlikely to suffer from the incompleteness due to the background filter. In the future, combination of the ALMA data with single-dish measurement can be used to address this caveat. We also do not assess in this work whether the column density peaks we identify correspond to peaks in local volume density. It is possible that some identified objects are column density peaks resulting from, e.g., overlap of unrelated structures along the line of sight. Further molecular line studies should be able to address this caveat in the future.


We give rough estimates for the masses of the ALMA-detected cores using the equation 
\begin{equation}
M = \frac{F_\nu d^2}{B_\nu (T)\kappa_\nu},
\label{eq:mass}
\end{equation}
where $F_\nu$ is the total integrated flux, $d$ is the distance, $B_\nu (T)$ is the Planck function at the temperature of $T$, and $\kappa_\nu$ is the mass absorption coefficient. For the temperature, we adopt 20 K (see \citealt{li13} for gas temperature measurements from NH$_3$). We calculate the mass absorption coefficient from the expression $\kappa_\nu = 0.1 (\nu / 1000 \ \mathrm{GHz})^\beta$ cm$^{2}$ g$^{-1}$, where $\beta$ is taken to be 1.5 \citep[e.g.,][]{sad16}. This expression assumes the gas-to-dust ratio of 100. The resulting mass absorption coefficient is $\kappa_{100 GHz} = 0.0032$ cm$^{2}$ g$^{-1}$. The masses of the cores span 0.2-2.6 M$_\odot$ (Table \ref{table:cores}). 


\begin{table*}
\centering
\caption{3 mm compact sources identified from the ALMA data.}
\begin{tabular}{lcccccclc}
\hline\hline
\# 		& $R.A.$ (2000) & $Dec.$ (2000) & $I_\mathrm{\nu}$ (peak)	&	$F_\nu$ 	& Size\tablefootmark{a} & $M$		&	Association\tablefootmark{b}		 &  class.\tablefootmark{c}   \\
		&			&		&	[mJy beam$^{-1}$]	&     [mJy]	&   & [M$_\odot$]	& &		\\
\hline
\multicolumn{9}{l}{OMC-3 / MMS 7} \\   
\hline
 1    & 05:35:28.165 & -05:03:41.25 &  13.6 &   8.3 &     4$\farcs$1$\times$2$\farcs$4 &   0.4 &     HOPS85  & ps(flat)\\
  2    & 05:35:26.541 & -05:03:54.91 &  49.4 &  39.2 &     4$\farcs$4$\times$2$\farcs$5 &   1.9 &     HOPS84 & ps(I)  \\
 \hline
\multicolumn{9}{l}{OMC-3 / MMS 8-10} \\     
\hline
   3    & 05:35:27.999 & -05:04:57.36 &  17.6 &   7.5 &     3$\farcs$9$\times$2$\farcs$5 &   0.4 &     HOPS81 & ps(0) \\
  4    & 05:35:25.950 & -05:05:43.24 &  40.3 &  42.2 &     2$\farcs$6$\times$4$\farcs$0 &   2.1 &     HOPS78 & ps(0) \\
  5    & 05:35:26.120 & -05:05:45.79 &   4.8 &   8.6 &     6$\farcs$2$\times$2$\farcs$6 &   0.4 &       -  & starless \\
  6    & 05:35:26.666 & -05:06:10.60 &   4.6 &   6.3 &     3$\farcs$8$\times$2$\farcs$4 &   0.3 &     HOPS75  & ps(0) \\
  7    & 05:35:24.849 & -05:06:21.63 &   5.7 &   3.9 &     4$\farcs$3$\times$2$\farcs$5 &   0.2 &     HOPS74  & ps(flat) \\
 \hline
\multicolumn{9}{l}{OMC-2 / FIR 1-2} \\   
\hline
  8    & 05:35:21.870 & -05:07:01.64 &   8.4 &   4.6 &     4$\farcs$8$\times$2$\farcs$6 &   0.2 &       -  & ps\tablefootmark{d} \\
  9    & 05:35:27.731 & -05:07:03.73 &   5.0 &  12.0 &     5$\farcs$4$\times$3$\farcs$5 &   0.6 &     HOPS73  & ps(0)\\
 10    & 05:35:23.507 & -05:07:09.79 &  18.3 &  17.0 &     5$\farcs$1$\times$2$\farcs$5 &   0.8 &       -  & starless \\
 11    & 05:35:24.188 & -05:07:52.75 &   9.7 &  38.1 &     6$\farcs$4$\times$4$\farcs$0 &   1.9 &       - & starless \\
 12    & 05:35:24.847 & -05:07:53.72 &  15.8 &  52.1 &     6$\farcs$3$\times$3$\farcs$0 &   2.6 &       019003 & PBRS\tablefootmark{e} \\
 13    & 05:35:25.584 & -05:07:57.34 &  33.3 &  15.2 &     3$\farcs$8$\times$2$\farcs$4 &   0.7 &     HOPS71 & ps(I) \\
 14    & 05:35:22.502 & -05:08:04.60 &   4.2 &   5.1 &     5$\farcs$9$\times$2$\farcs$5 &   0.3 &     HOPS70 & ps(flat) \\
 15    & 05:35:24.009 & -05:08:27.65 &   4.2 &   8.0 &     6$\farcs$4$\times$3$\farcs$2 &   0.4 &       - & starless \\
 16    & 05:35:24.287 & -05:08:30.65 &  21.1 &  33.6 &     4$\farcs$2$\times$3$\farcs$2 &   1.7 &     HOPS68 & ps(I) \\
 \hline
\multicolumn{9}{l}{OMC-2 / FIR 3-5} \\   
\hline
17    & 05:35:26.928 & -05:09:24.40 &   8.2 &  30.4 &     5$\farcs$7$\times$4$\farcs$2 &   1.5 &     HOPS66, SOF1 & ps(flat) \\
 18    & 05:35:27.197 & -05:09:26.10 &   3.3 &   5.6 &     3$\farcs$7$\times$4$\farcs$4 &   0.3 &       - & starless \\
 19    & 05:35:27.512 & -05:09:29.54 &   3.3 &  13.2 &     4$\farcs$0$\times$6$\farcs$6 &   0.7 &       - & starless \\
 20    & 05:35:27.618 & -05:09:34.06 &  70.3 &  51.7 &     3$\farcs$7$\times$2$\farcs$6 &   2.5 &     HOPS370, & ps(I) \\
\multicolumn{7}{l}{ } & FIR3, SOF2N, &  \\
\multicolumn{7}{l}{ } & CO outflow &  \\
 21    & 05:35:21.566 & -05:09:38.62 &  12.3 &   4.7 &     3$\farcs$3$\times$2$\farcs$7 &   0.2 &     HOPS65 & ps(I) \\
 22    & 05:35:26.992 & -05:09:48.62 &   4.9 &   7.3 &     5$\farcs$3$\times$2$\farcs$9 &   0.4 &       - & starless \\
23    & 05:35:27.528 & -05:09:50.64 &   2.3 &  14.0 &     9$\farcs$2$\times$3$\farcs$9 &   0.7 &       - & starless \\
 24    & 05:35:26.928 & -05:09:54.51 &   6.4 &  10.1 &     2$\farcs$8$\times$4$\farcs$6 &   0.5 &       HOPS64 & ps\tablefootmark{d} \\
 25    & 05:35:27.015 & -05:09:59.59 &   3.9 &   5.9 &     5$\farcs$9$\times$2$\farcs$5 &   0.3 &     HOPS108, & ps(0) \\
\multicolumn{7}{l}{ } & FIR4, SOF3 &  \\
 26    & 05:35:26.797 & -05:10:04.81 &   4.5 &  11.6 &     3$\farcs$6$\times$5$\farcs$6 &   0.6 &       - & starless \\
 27    & 05:35:26.455 & -05:10:05.10 &  10.6 &  20.6 &     2$\farcs$8$\times$5$\farcs$3 &   1.0 &       - & starless \\
 28    & 05:35:26.396 & -05:10:15.20 &   4.5 &  37.3 &     5$\farcs$1$\times$9$\farcs$4 &   1.8 &       - & starless \\
 29    & 05:35:26.325 & -05:10:25.13 &   3.8 &  24.7 &     5$\farcs$0$\times$7$\farcs$6 &   1.2 &       - & starless \\
 30    & 05:35:24.717 & -05:10:29.78 &  27.4 &  15.2 &     3$\farcs$9$\times$2$\farcs$4 &   0.7 &     HOPS368, SOF5 & ps(I) \\
 \hline
\multicolumn{9}{l}{OMC-2 / FIR 6}\\    
\hline
 31    & 05:35:23.278 & -05:12:03.15 &  44.7 &  34.3 &     2$\farcs$4$\times$4$\farcs$0 &   1.7 &     HOPS60, FIR6b, & ps(0) \\
 \multicolumn{7}{l}{ } & CO outflow &  \\
 32    & 05:35:22.745 & -05:12:26.47 &   9.9 &  24.9 &     4$\farcs$0$\times$4$\farcs$5 &   1.2 &       FIR6a & starless \\
 33    & 05:35:22.939 & -05:12:40.70 &   2.7 &  14.8 &     7$\farcs$2$\times$4$\farcs$8 &   0.7 &       FIR6a & starless \\
 34    & 05:35:18.221 & -05:13:05.83 &  14.7 &   7.1 &     4$\farcs$1$\times$2$\farcs$6 &   0.4 &       - & ps\tablefootmark{d} \\
 35    & 05:35:21.705 & -05:13:13.08 &   3.1 &  12.7 &     4$\farcs$1$\times$7$\farcs$0 &   0.6 &       FIR6c & starless \\
 36    & 05:35:20.138 & -05:13:15.40 &  67.3 &  40.9 &     3$\farcs$8$\times$2$\farcs$4 &   2.0 &     HOPS59, FIR6d, & ps(flat) \\
\multicolumn{7}{l}{ } & CO outflow &  \\
 37    & 05:35:21.348 & -05:13:17.54 &  27.9 &  22.5 &     4$\farcs$2$\times$2$\farcs$4 &   1.1 &       HOPS409, FIR6c,  & ps(0) \\
\multicolumn{7}{l}{ } & CO outflow &  \\
 38    & 05:35:20.760 & -05:13:22.51 &   4.6 &   9.1 &     5$\farcs$3$\times$3$\farcs$0 &   0.4 &       FIR6c & starless \\
 39    & 05:35:18.516 & -05:13:38.20 &  34.9 &   9.5 &     3$\farcs$9$\times$2$\farcs$5 &   0.5 &     HOPS58 & ps(flat) \\
\hline
 40    & 05:35:19.872 & -05:15:08.02 &  27.1 &  16.2 &     3$\farcs$9$\times$2$\farcs$4 &   0.8 &     HOPS57 & ps(flat) \\
 41    & 05:35:19.503 & -05:15:32.70 &   4.1 &   9.3 &     4$\farcs$6$\times$3$\farcs$9 &   0.5 &     HOPS56 & ps(0) \\
 42    & 05:35:25.265 & -05:15:35.25 &  31.0 &  17.7 &     4$\farcs$1$\times$2$\farcs$4 &   0.9 &       - & ps\tablefootmark{d} \\
 43    & 05:35:19.413 & -05:15:37.98 &   9.5 &  11.0 &     2$\farcs$6$\times$4$\farcs$3 &   0.5 &       - & starless \\
\end{tabular}
\tablefoot{
\tablefoottext{a}{The $FWHM$ sizes of the object.}
\tablefoottext{b}{Association with the far-infrared sources as labelled in \citep{mez90, chi97}, HOPS protostars \citep{fur16}, PBRSs \citep{stu13}, SOFIA mid-infrared detections \citep{ada12}, and CO outflows \citep{shi08, shi09}.}
\tablefoottext{c}{Evolutionary classification of the object. Classification of protostars (ps) is according to \citet{fur16} unless otherwise stated. The objects without identified protostars are considered starless.}
\tablefoottext{d}{The core coincides with a disk source \citep{meg12}. Due to the association with extended 3 mm emission and the low expected number of chance superpositions, we re-classify the source as a protostar.}
\tablefoottext{e}{PBRS, i.e., a Class 0 source with a very red spectral energy distribution \citep{stu13}.}
}
\label{table:cores}
\end{table*}

\section{Results} 
\label{sec:results}

\subsection{Structure of the OMC-2/3 at 1\,000 AU resolution}
\label{subsec:maps}

\begin{figure*}
   \centering
\includegraphics[bb = 170 10 1460 650, clip=true, angle=90, width=0.63\textwidth]{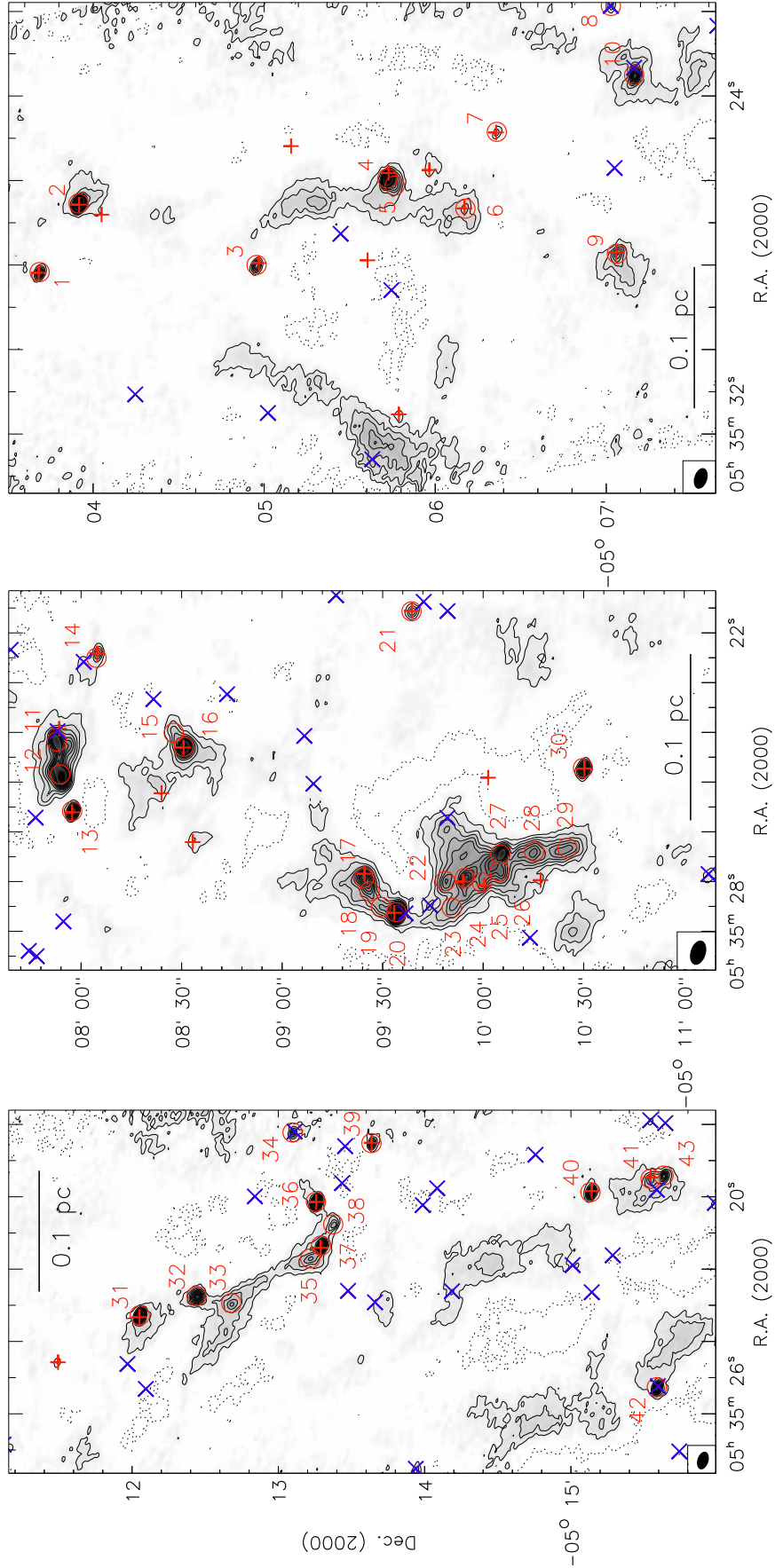}
      \caption{ALMA 3 mm continuum emission maps of the OMC-2/3 region (combined 12-m and 7-m array data). The beamsize is $3\farcs75 \times 2\farcs27$ and the $FWHM$ beam is shown in the lower left corner of the frames. The contours are drawn in $3\sigma_\mathrm{rms}$ intervals, with $\sigma_\mathrm{rms}=0.23$ mJy beam$^{-1}$. The dotted contour is at $-3\sigma_\mathrm{rms}$. The red circles and numbers indicate the objects detected by \textsf{gaussclumps} (listed in Table \ref{table:cores}). These objects are regarded as dense cores, unless they coincide with a star with disk. The red plusses indicate protostars and the blue crosses stars with disks. 
              }
         \label{fig:results3}
   \end{figure*}

The ALMA map reveals intricate details within the OMC-2/3 region (Fig. \ref{fig:results3}). Most structures identified in the previous studies at $\sim$10$\arcsec$ resolution \citep{mez90, chi97, joh99}, and in the \emph{Herschel} data at $\sim$18$\arcsec$ resolution \citep[Fig. \ref{fig:footprint};][]{stu15} are resolved into multiple sub-structures. Some of the sub-structures are filamentary. In the following, we briefly describe the morphology as revealed by the ALMA map beyond the previous studies. We follow the designations of \citet{mez90} and \citet{chi97} (shown in Fig. \ref{fig:results}), proceeding from north to south.

\emph{MMS 7} (core \#2): The source is single peaked at the resolution of the ALMA map (as already indicated by the $4\arcsec$ resolution data of \citealt{tak13} at 1300 $\mu$m). We detect a previously undetected compact source $14\arcsec$ (5\,800 AU) North-East from MMS 7 (core \#1).

\emph{MMS 8-9} (cores \#4-7): The prominent filamentary structure detected by \citet{mez90} and \citet{chi97} breaks into three local maxima, north- and centre-most of which correspond to MMS 8 and MMS 9, respectively.  

\emph{MMS10} (no ALMA cores): The appearance is similar to that evidenced by the \citet{chi97} data, with one local maximum in an elongated clump. 
 
 \emph{FIR 1a, b, c} (cores \#8, 10-14): The ALMA data resolves FIR 1a into four dense cores \citep[three of them have previously been detected from arcsecond-resolution continuum emission data by][]{tob15}. One of the peaks (core \#11) is close ($3\farcs5$, or $\sim$1\,500 AU) to a Class 0 protostar that has exceptionally high envelope density and mass-infall rate, discovered by \citet{stu13} (referred to as the \emph{PACS} Bright Red Sources, PBRSs). The configuration of the two strongest peaks (cores \#11 and \#12) bears resemblance to the Bok Globule \object{CB 244} in which a starless core and a young protostar co-exist within a common envelope \citep{stu10}. FIR 1b is relatively extended in the ALMA map and does not harbour a dense core. FIR 1c contains a single compact source. 
 
 \emph{FIR 2} (cores \#15-16): Two cores are found close to the peak identified by \citet{chi97}.
 
\emph{FIR 3-5} (cores \#17-20, 22-30): \citet{chi97} identifies three sources in the region from mm-wavelength data. Mid-infrared imaging at $\sim$$4\arcsec$ resolution resolves eight sources in the region \citep{ada12} and mm-wavelength data detect 12 fragments altogether \citep{shi08}. The ALMA data shows how the region is inter-connected by a filamentary structure from which we identify 13 dense cores. Four cores coincide with HOPS protostars. The cores \#20 and 25 coincide with the FIR 3 and 4 \citep[see, e.g.,][]{fur14}. FIR 5 does not appear as a dense core in the ALMA data, although, it is within $\sim$4\arcsec of the cores \#28 and 29. Previously, \citet{shi08} suggested that the fragmentation in the area is caused by interaction with the outflow driven by FIR 3 with a nearby clump. In general, we find a rather poor correspondence between the ALMA-detected dense cores and the fragments identified by \citet{shi08}. The ALMA data indicate that the cores are primarily arranged along a filament in this region.

\emph{FIR 6a-e} (cores \#31-33, 35-38): Another prominent filamentary structure in the ALMA data. The sources FIR 6a-d identified by \citet{chi97} are each resolved into at least two dense cores. The source FIR 6e is filamentary in the ALMA data and does not contain a compact source. Altogether, the region contains nine dense cores. The south-most tip of the structure harbours four sources within a $\sim$12\,000 AU region, making it the second densest concentration of sources in the mapped area after the FIR 3-5 filament. 

Our ALMA map continues about $2\arcmin$ south from the area mapped by \citet{chi97}. This area contains two prominent structures elongated in north-south direction. The western of these is clearly visible in the \emph{Herschel}-derived column density map \citep[][see Fig. \ref{fig:footprint}]{stu15}. The eastern structure is not visible in the \emph{Herschel} data, but is clearly visible in the 850 $\mu$m emission data \citep[][also partly visible in the data of \citealt{joh99}]{lan16} and also detected in ammonia line emission (J. Pineda, priv. communication). The area contains four dense cores (\#40-43), listed in Table \ref{table:cores}.


\subsection{Nearest neighbour separations of the dense cores, protostars, and disks}
\label{subsec:separations}


Our goal is to characterise the fragmentation in the ISF down to $\sim$1\,000 AU scales with a special emphasis in searching for preferential fragmentation scales, and to subsequently compare the characteristics to those of the young stellar population of the filament. First, we compute the basic statistics of the projected nearest neighbour separations of the objects in the ALMA-covered region. Note that the sample of 43 dense cores consists of 17 starless cores and 26 protostellar cores. However, there are altogether 37 protostars in the mapped area \citep[33 from][and four re-classified by us form \citealt{meg12}]{fur16}; not all protostars coincide with an object identified as a core, which may indicate that some HOPS protostars may be misclassified stars with disks. We report the mean and median nearest neighbour separation, $d_\mathrm{nn}$, for all different object classes in Table \ref{tab:objects}. The mean projected nearest neighbour separations of the dense cores are similar to those seen in the OMC-3 region of the ISF by \citet{tak13}. They are a factor of two larger than those measured in the ONC-1n region by \citet{tei15}. The mean nearest neighbour separations of protostars and stars with disks are roughly a factor of two smaller than in Orion A on average \citep{meg16}. However, we show below and in Sect. \ref{subsec:2point} that the cores, protostars, and disks are clearly grouped in different ways. This implies that the mean nearest neighbour separations may not be very useful measures, especially when averaged over a variety of environments. 

\begin{figure}
   \centering
   \includegraphics[bb = 35 5 490 335, clip=true, width=\columnwidth]{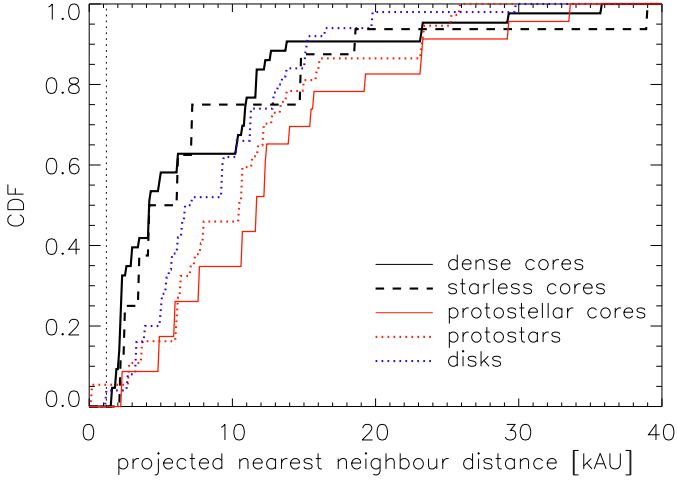}
      \caption{Cumulative distribution function, CDF, of the projected nearest neighbour separations for all dense cores (solid black line), starless cores (dashed black line), protostellar cores (red line), protostars (dotted red line), and stars with disks (dotted blue line). The vertical dotted line shows the 1\,200 AU resolution limit of the ALMA data.  
              }
         \label{fig:cdf}
   \end{figure}

Figure \ref{fig:cdf} shows the cumulative distribution functions (CDFs) of the nearest neighbour separations of all cores, starless cores, protostars, and stars with disks. The CDFs show that the nearest neighbour distribution of starless cores and protostellar cores are different, with the starless cores exhibiting a larger fraction of short separations. The short core separations increase down to the resolution limit of the ALMA data (1\,200 AU). Thus, we do not detect a preferential nearest-neighbour separation for the dense cores. Note here the caveat that our starless core sample may be incomplete and may include objects that are enhancements only in column density, not in volume density (see Section \ref{subsec:core-detection}). The CDFs also show that the distribution of the dense cores is different from that of protostars and disks, specifically so that cores have an excess of short separations compared to disks and protostars. 


\begin{table}
\caption{Properties of the nearest neighbour distributions.}             
\label{tab:objects}     
\centering                    
\begin{tabular}{l c c c c}     
\hline\hline            
		Objects	& No.&	$<d_\mathrm{nn}>$   	& median $d_\mathrm{nn}$   & $<M>$	\\
				&		&	[kAU]		& 	[kAU]		&	 [M$_\odot$] \\	
\hline
All dense cores			& 43			& 7.7	&	 4.2		& 		0.9 \\
Starless cores			& 17			& 11.5&    6.1	&		0.8 \\
Protostellar cores		& 26			& 12.9&	11.7	&		1.0 \\
Protostars 			& 37			& 11.4 		&	10.6		&		-		\\
Stars with disks   		& 51   			&	  8.7	& 7.1			&		-		\\
\hline
\end{tabular}
\end{table}

\subsection{Two-point correlation function of the dense core, protostar, and disk separations}
\label{subsec:2point}


To analyse the grouping of the ALMA cores, we examine the two-point correlation function, $\xi (r)$, of the core separations. The function describes the excess probability per unit area, in comparison to the expectation from random placement, of an object being located at a separation $r$ from another object. We adopt the \citet{lan93} estimator for the two-point correlation function
\begin{equation}
\xi (r) = \frac{\frac{2}{N_\mathrm{cores}(N_\mathrm{cores}-1)}DD(r) - \frac{2}{N_\mathrm{cores}^2}DR(r) + {\frac{2}{N_\mathrm{cores}(N_\mathrm{cores}-1)}}RR(r)}{\frac{2}{N_\mathrm{cores}(N_\mathrm{cores}-1)}RR(r)},
\label{eq:2point}
\end{equation} 
where DD($r$) is the observed distribution of pair separations (for $N_\mathrm{cores}$ cores), DR($r$) is the distribution of pair separations when one core is taken from the observed sample and one from a random sample (resulting in $N_\mathrm{cores}^2$ pairs), and RR($r$) is the distribution of pair separations of randomly placed cores (for $N_\mathrm{cores}$ cores). The $\xi (r) = 0$ indicates a distribution similar to random distribution; the positive and negative values indicate excess and deficit of separations, respectively. We compute the random separation distributions (RR($r$)) with a Monte Carlo simulation, in which we place $N_\mathrm{cores}=43$ points in the ALMA map area and compute their pair separation distribution. This is repeated 10\,000 times (and the resulting distribution is divided by 10\,0000) to obtain adequate statistics. The cross term (DR($r$)) is computed via a similar simulation, only with one point taken from among the observed core locations and another from among random locations. We calculate all separation distributions using a Gaussian density estimator that has a $FWHM=5\arcsec$ (2\,100 AU). For example, DD($r$) then describes the separation distribution within 2\,100 AU bin centered at $r$. We tested whether the choice of the $FWHM$ affects the main conclusions we derive using two-point correlation functions significantly, and it does not.   

\begin{figure}
   \centering
\includegraphics[width=\columnwidth]{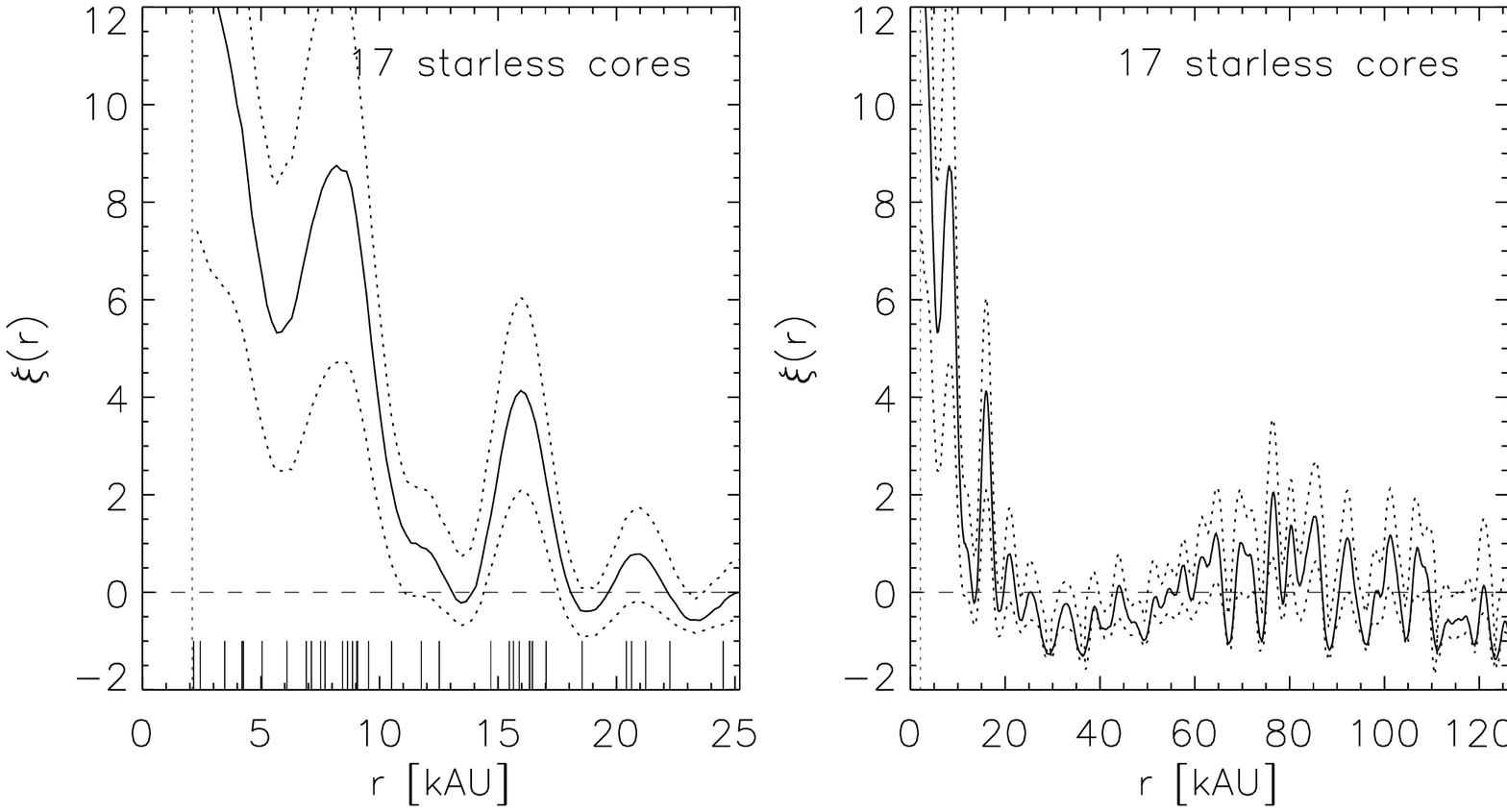}
\includegraphics[width=\columnwidth]{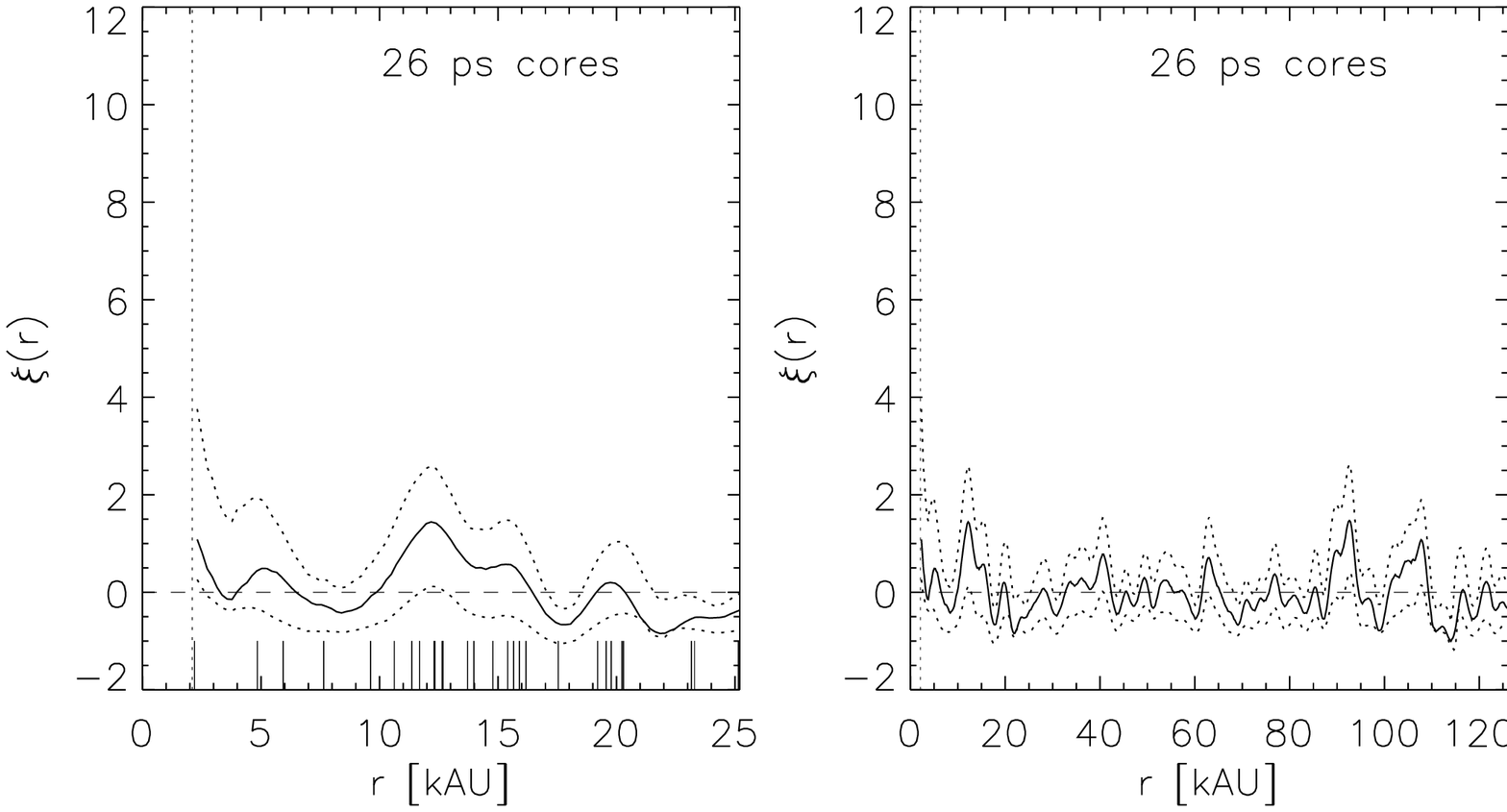}
\includegraphics[width=\columnwidth]{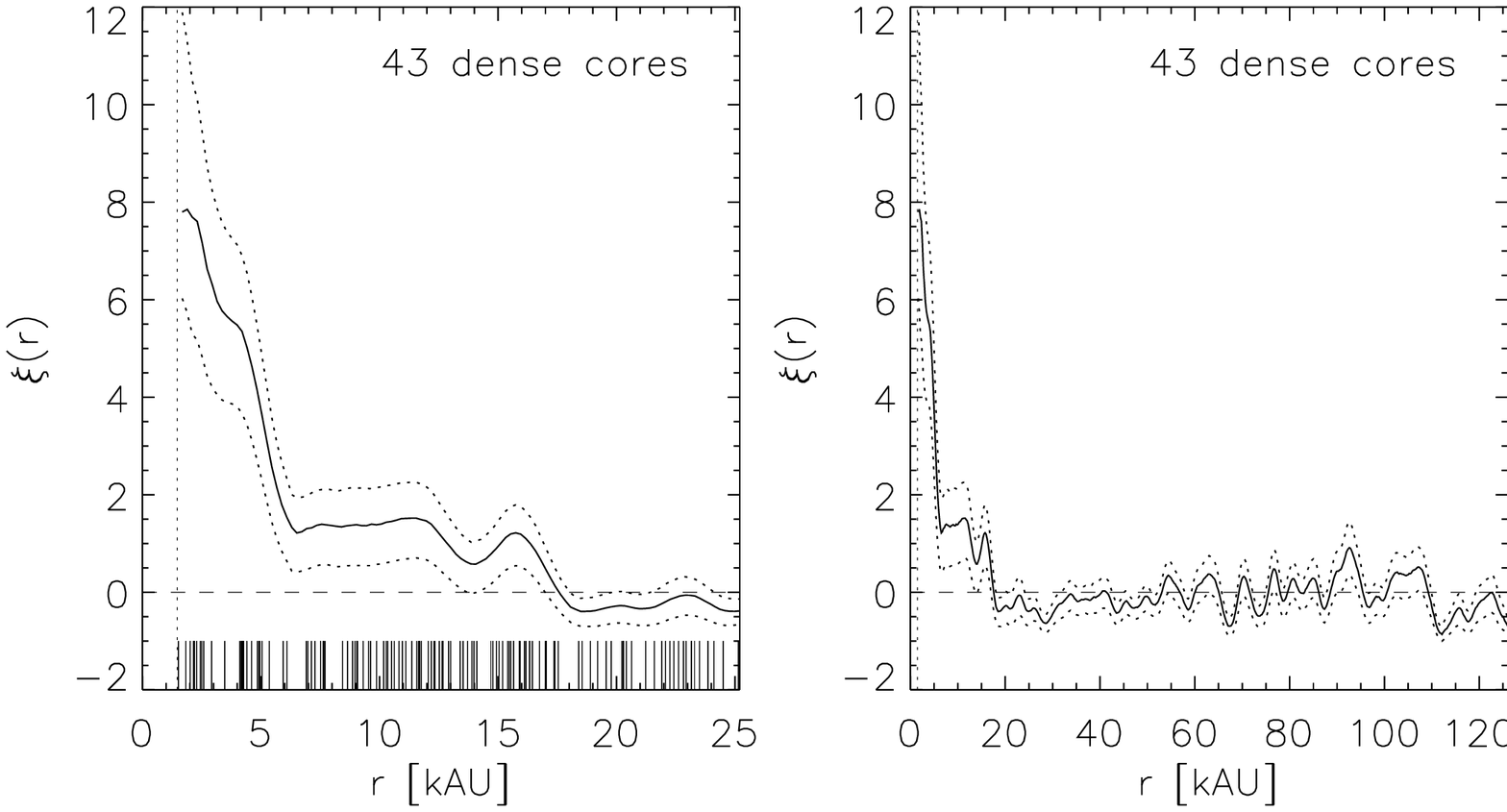}
      \caption{Two-point correlation function of the separations of all ALMA dense cores, starless cores, and protostellar cores identified from the ALMA data. The frames on the left show the separations until 25 kAU and the frames on the right until 125 kAU. The dotted curves show the confidence intervals given by three times the 25\% and 75\% quartiles. The vertical dotted line shows the minimum separation. The dashed horizontal line, drawn at zero, indicates a random distribution. The short vertical lines at the bottom of the panels show the observed separations (only in the panels on the left). 
              }
         \label{fig:2point-cores}
   \end{figure}

\begin{figure}
   \centering
\includegraphics[width=\columnwidth]{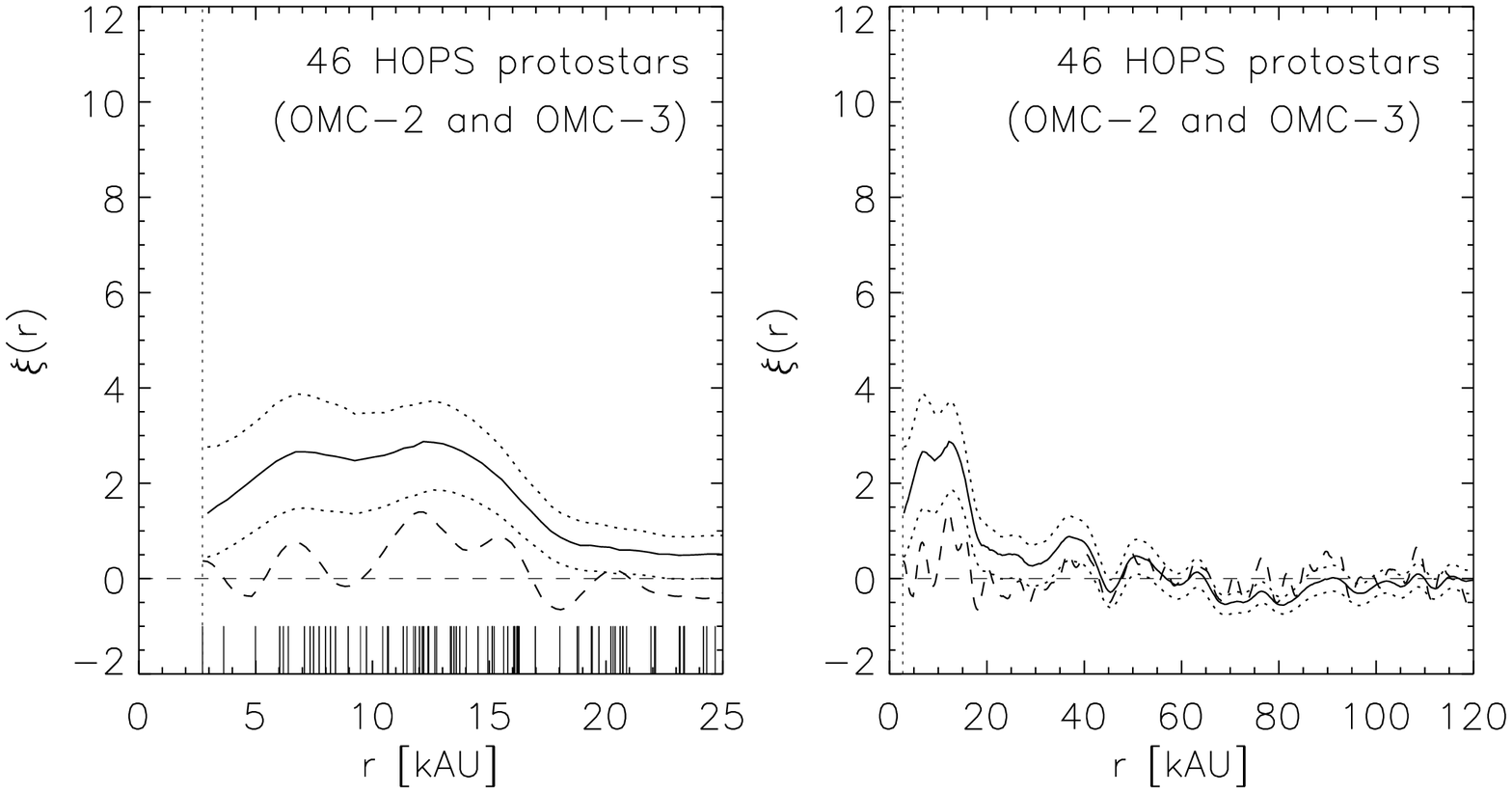}
\includegraphics[width=\columnwidth]{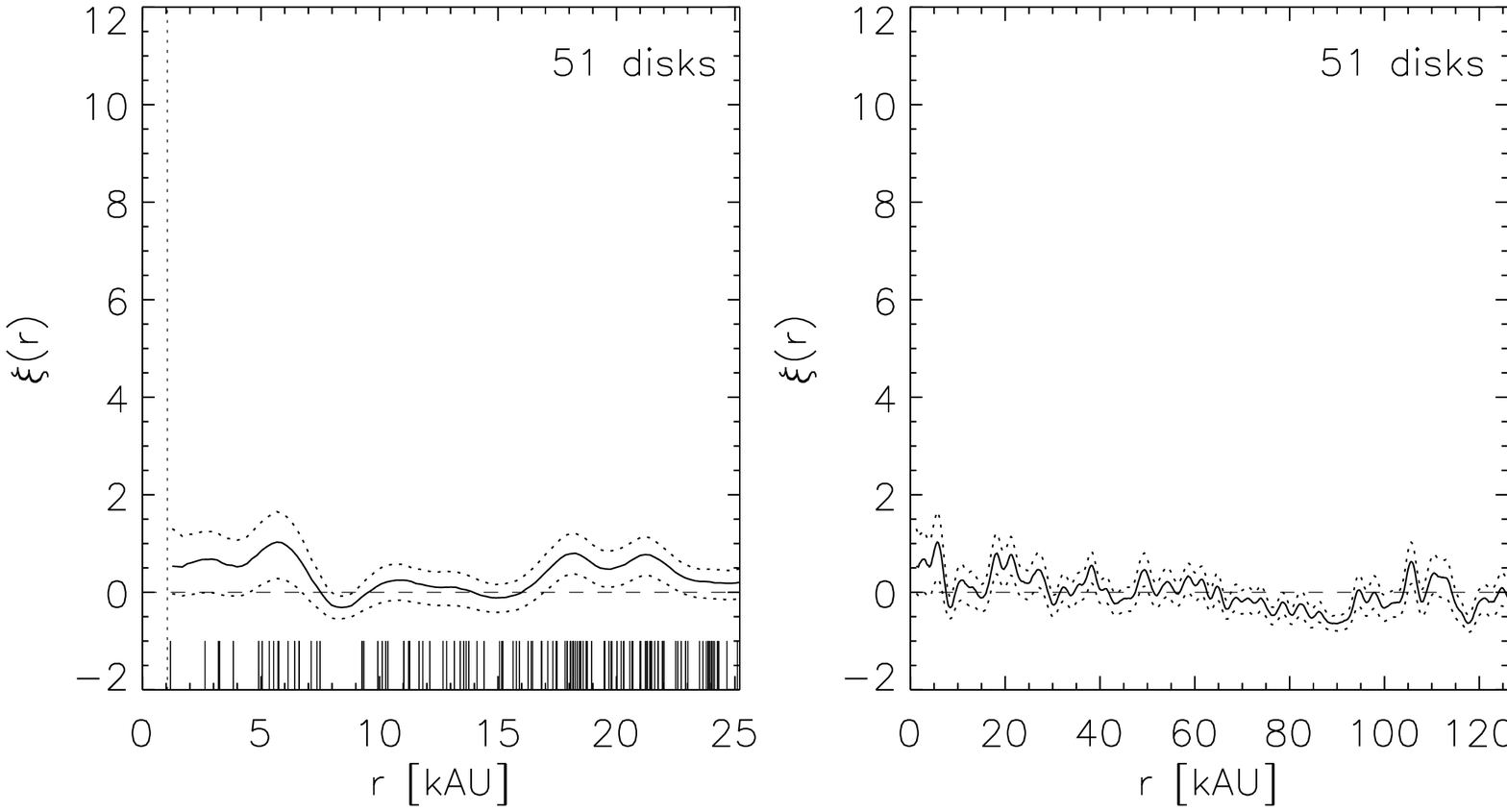}
      \caption{Two-point correlation function of the separations of the protostars and stars with disks. The frames on the left show the separations until 25 kAU and the frames on the right until 125 kAU. The dotted curves show the confidence intervals given by three times the 25\% and 75\% quartiles. The vertical dotted line shows the minimum separation. The dashed horizontal line, drawn at zero, indicates a random distribution. The short vertical lines at the bottom of the panels show the observed separations (only in the panels on the left). The frames for the protostars show the function combining the protostars from the OMC-2 and 3 regions with a solid line. The function obtained for the region mapped by ALMA is shown with a dashed line (for visibility, the confidence intervals are not shown). 
              }
         \label{fig:2point-hops}
   \end{figure}

We compute an uncertainty estimate of the two-point correlation function using bootstrapping. We randomly draw $N_\mathrm{cores}$ points from all observed core locations and calculate their separation distribution, resulting in a simulated sample and the corresponding $DD(r)$ function. The two-point correlation function is then computed following the steps described above.
The procedure is repeated 10\,000 times, resulting in a sample of 10\,000 $\xi(r)$ functions. For each $r$, we compute the 25\% and 75\% quartiles of these functions and use the quartiles as uncertainty estimates (shown in Figs. \ref{fig:2point-cores} and \ref{fig:2point-hops}).


The two-point correlation functions of the 17 starless cores, 26 protostellar cores, and all 43 cores together are shown in Fig. \ref{fig:2point-cores}. All are close to zero above 17\,000 AU (however, see Sect. \ref{sec:discussion}). The two-point correlation function of the starless cores is dramatically different from that of the protostellar cores. The starless cores show a systematic increase of grouping at short separations down to our resolution limit. The protostellar cores are in agreement with a random distribution practically at all separations. However, one has to recognise that the number of detections especially below $\sim$10\,000 AU is very small, and consequently, the uncertainty is large. The two-point correlation function of all 43 cores together is sampled reasonably well down to our resolution limit, and it shows significant grouping below 17\,000 AU and strong grouping below $\sim$6\,000 AU. These results are not strongly affected by possible misclassification of core-like objects (see Appendix \ref{app:robustness}). 


We additionally analyse the two-point correlation function of the protostars (Fig. \ref{fig:2point-hops}). For the 37 protostars in the ALMA-covered area, the function is in agreement with a random distribution. However, there are very few short ($\lesssim$10\,000 AU) separations, hampering the accuracy of the function; the function is also in agreement with a significant excess of short separations. To better quantify the distribution of protostars in the ISF area, we calculated the two-point correlation function for all 46 protostars in the northern ISF area (see Fig. \ref{fig:footprint}). For this higher number of objects, the uncertainties are smaller and the function indicates a significant excess of separations shorter than $\sim$17\,000 AU and a marginal excess at separations shorter than roughly $\sim$40\,000 AU. This is somewhat in contradiction with the earlier result that the protostellar cores are in agreement with random distribution. This may result from the small number of protostellar cores in the ALMA-covered region that prohibits detecting a marginal excess of separations. In summary, the data suggest that the dense cores and protostars in the northern ISF show a similar excess of separations between  $\sim$6\,000 - 17\,000 AU. 


The two-point correlation function of the 51 stars with disks within the ALMA-covered region is consistent with zero at almost all separations (there is a marginally significant positive peak at $\sim$5\,000 AU, see Fig. \ref{fig:2point-hops}). This indicates that the separation distribution of disks in the region is in agreement with a random distribution, and therefore, it is significantly different from the distributions of the ALMA dense cores and protostars in the northern ISF. 


 \section{Discussion}
\label{sec:discussion}

Here, we first discuss the fragmentation picture implied by the observed separation statistics of the different object classes (Sect. \ref{subsec:picture}). Then, we discuss the picture in the context of theoretical gravitational fragmentation models (Sect. \ref{subsec:jeans}).

 \subsection{The fragmentation picture implied by the spatial distribution of the cores, protostars, and disks}
\label{subsec:picture}


Most crucially, we detect \emph{increasing grouping} of ALMA dense cores at separations below $\sim$17\,000 AU, and especially below $\sim$6\,000 AU. We also show that the frequency of short nearest neighbour separations of the cores increases down to our resolution limit (1\,200 AU); in other words, we do not detect any peaks in the nearest neighbour separation indicating well-defined preferential separations. These observations directly point towards hierarchical, scale-dependent fragmentation of the ISF. The continuous increase of short separations is especially interesting for the models of multiplicity during star formation; numerical simulations predict that one efficient formation mechanism of bound multiple systems is core fragmentation at scales between 500-5\,000 AU \citep[e.g.,][]{off10}. Especially, pairs formed at separations below roughly 3\,000-4\,000 AU are likely to result in a gravitationally bound multiple system. The strong increase of short core separations we detect below $\sim$6\,000 AU may mark the regime of these systems \citep[see, e.g.,][for a discovery of a bound core system at these scales]{pin15}. 

Importantly, the strong grouping of cores below 6\,000 AU is driven by the starless cores, while the protostellar cores do not show such a grouping signature (cf., Fig. \ref{fig:2point-cores}). One should keep in mind the caveats regarding the core identification that may be significant especially for the starless cores (cf., Sect. \ref{subsec:core-detection}). The majority of the starless cores are located close to ($\lesssim$ 5\,000 AU) protostellar cores, indicating a possibility that some may be column density enhancements caused by, e.g., outflow cavities. However, the effect of possible misclassifications to the two-point correlation analysis is likely not strong (see Appendix \ref{app:robustness}). With these caveats in mind, we speculate about the possible origin of the grouping difference between the protostellar and starless cores. One possibility is an evolutionary sequence in which the protostellar cores have lost some of their grouping inherited from the fragmentation process, but this signature is still imprinted in the grouping of starless cores. Other hypotheses could be that not all cores end up forming stars or that the cores merge during their evolution. Further studies on the kinematics of the ALMA-detected cores could address the strength of the increase in core separations below $\sim$6\,000 AU and explore the origin of it.


Previously, \citet{tei15} have studied a section of the ISF in \object{OMC-1n} and claimed evidence of two preferred nearest-neighbour core separation scales, at 5\,500 AU and 13\,000 AU. 
Our data does not indicate peaks in the preferred core separations, but rather a continuous increase of short separations. This difference may be due to various reasons. For example, \citet{tei15} only had a sample of 24 cores, hampering the operation of statistical estimators; they do not present an analysis of how significant their detections of the preferred scales are. Also, our core identification method is different from that of \citet{tei15}, making a direct comparison difficult. 
%
%
Studies targeting other massive filaments have commonly quantified mean separations between structures detected at different spatial scales \citep[e.g.,][]{jac10, mie12, kai13, wan14, beu15a, con16}. However, none of these works to our knowledge has analysed the higher-order statistics of the fragment separations and linked them to the young stellar distribution in the clouds. 

\begin{figure}
   \centering
\includegraphics[bb = 5 30 410 655, clip = true, width=0.85\columnwidth]{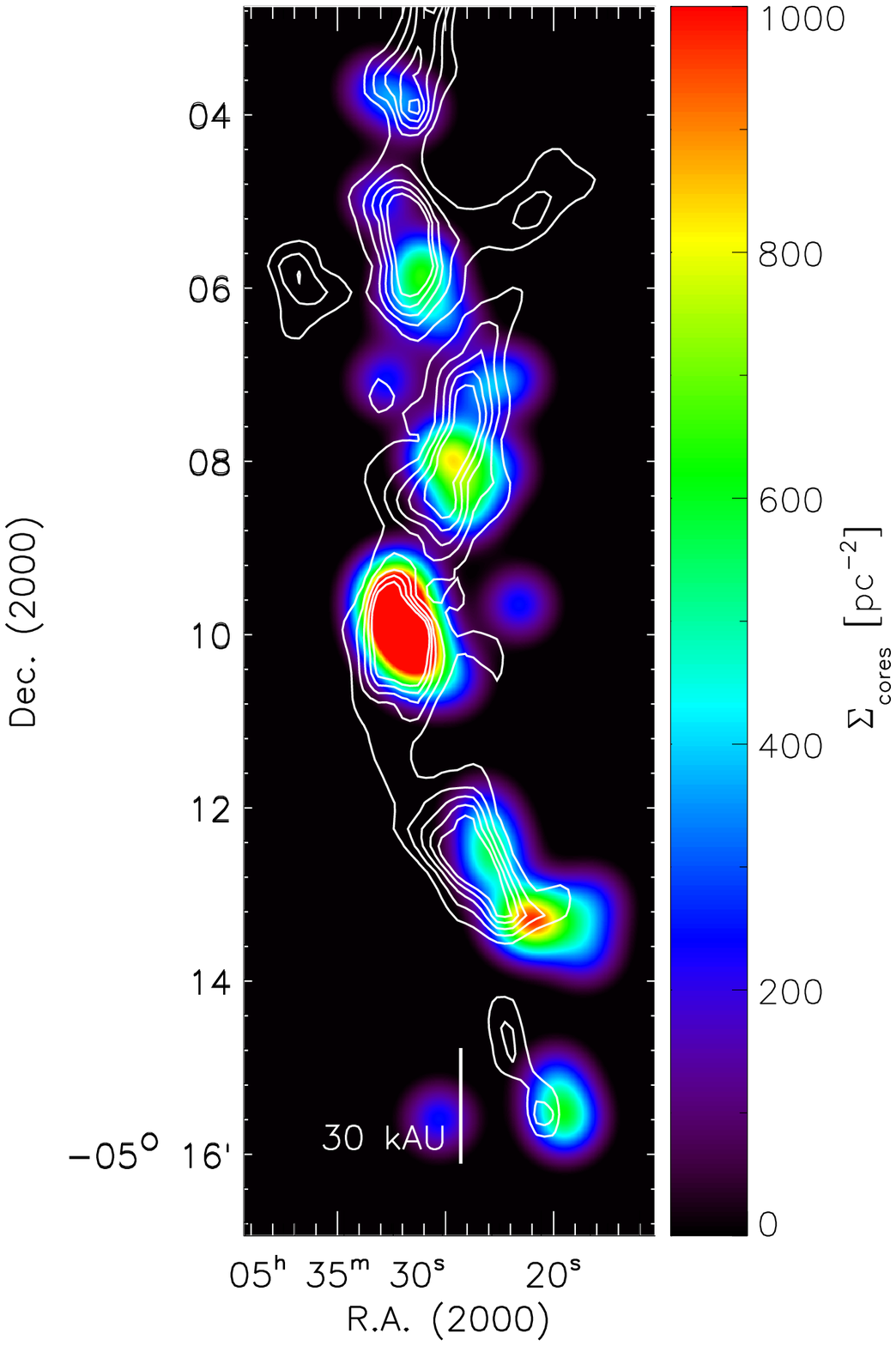}
\includegraphics[bb = 0 5 320 330, clip = true, width=0.75\columnwidth]{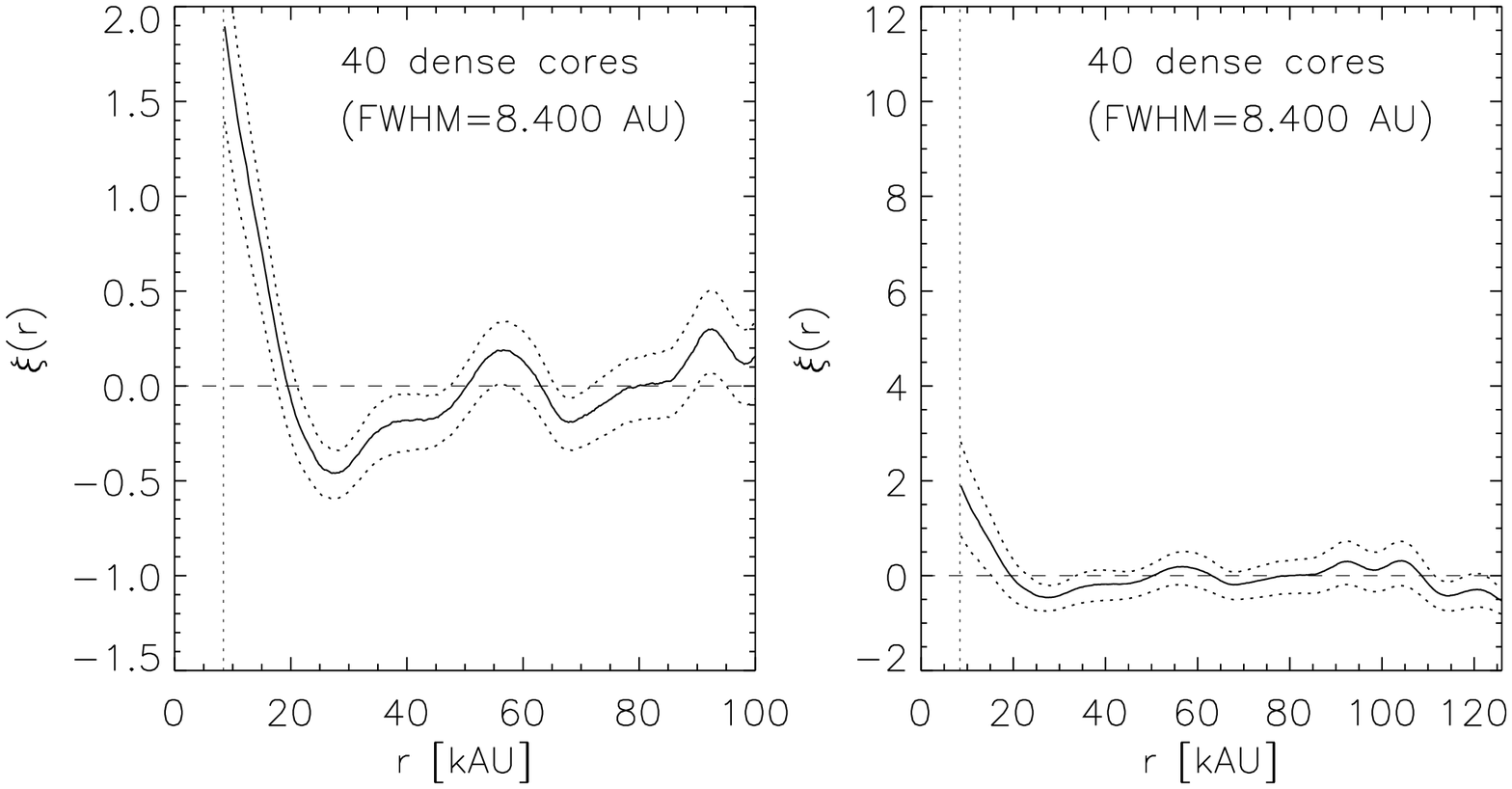}      \caption{\emph{Top: }the surface number density of the dense cores identified from the 3 mm ALMA continuum data. The white contours show the \emph{Herschel}-derived column densities at $N$(H)=$\{60, 80, 100, 120, 140\} \times 10^{21}$ cm$^{-2}$ \citep{stu15}. \emph{Bottom left: }the two-point correlation function of the 43 dense cores, derived with a smoothing function with $FWHM=20\arcsec = 8.400$ AU.
              }
         \label{fig:2point-sim}
   \end{figure}

Our analysis of the core distribution has so far concentrated on the small scales. We now consider the distribution at the scales larger than the grouping identified by the two-point correlation analysis ($>17\, 000$ AU). Figure \ref{fig:2point-sim} shows the number of detected ALMA cores per unit area, derived using a Gaussian smoothing kernel that has the $FWHM$ of 40$\arcsec$ (17\, 000 AU). The distribution is clearly not random, but concentrated on groups along the filament. The groups are roughly 30\,000 AU in size and they are separated quasi-periodically by roughly 50\,000 AU (62\,000 $\pm$ 19\,000 AU when considering the six largest groups and 42\,000 $\pm$ 26\,000 AU if considering the feature at around Dec. $\approx$ -5\fdg13\fm00 as two groups). The core groups coincide with the local maxima of dust column density as probed by \emph{Herschel} that have the mean separation of 56\,000 $\pm$ 14\,000 AU (see Fig. \ref{fig:2point-sim}). Thus, the ALMA and \emph{Herschel} data provide independent measurements of the quasi-periodic fragmentation mode and they indicate that the neighbouring cores have common, tenth-of-a-parsec-scale envelopes.

The quasi-periodic fragmentation mode can also be inferred from the two-point correlation function of the cores ALMA. We looked for signals in the function at large scales by heavily smoothing it to gain a better signal-to-noise. The function derived with a smoothing function that has $FWHM$=20$\arcsec$=8\,400 AU is shown in Fig. \ref{fig:2point-sim}. The function shows a significant negative feature between $\sim$20\,000-40\,000 AU, indicating a deficit of cores at these separations with respect to a random distribution (note that by definition, the two-point correlation function cannot have values lower than -1, see Eq. \ref{eq:2point}). The negative feature is followed by a marginally significant peak at $\sim$55\,000 AU. We show in Appendix \ref{app:2point} that marginally significant negative and positive signatures are expected for periodically grouped cores at the characteristic size of the groups and at the wavelength of the grouping period, respectively.
 

In summary, the spatial distribution of the ALMA-detected cores indicates periodic grouping into core groups roughly 30\,000 AU in size, separated by roughly 55\, 000 AU. At scales smaller than 30\,000 AU, i.e., inside the individual groups, the cores group below about 17\, 000 AU and especially strongly below 6\,000 AU (cf., Fig. \ref{fig:2point-cores}). 

The above picture is in agreement with previous studies that have found signatures interpreted as co-existing spherical and cylindrical gravitational fragmentation modes in highly super-critical filaments, with the transition between the two occurring at a few tenths of a parsec \citep[0.3 pc = 60\,000 AU;][]{kai13, tak13}. These works have suggested that the transition could result, for example, from the finite-size effects in filaments: below a certain size-scale local collapse may proceed faster than the global collapse of the filament \citep[e.g.,][]{pon11, kai13}. Regardless of its driver, our data suggests that a similar signature is detectable in the separation distribution of the cores: the 55\,000 AU separations between the groups correspond to the scale of the filamentary fragmentation mode, and the increasing separations below $\sim$17\,000 AU correspond to the scale of the spherical fragmentation mode. We further address the feasibility of this interpretation in Sect. \ref{subsec:jeans}.


The relationship between the separation distributions of the ALMA dense cores and HOPS protostars and disks can indicate temporal evolution. The dense core and protostar separation distributions show similarity between $\sim$6\,000 - 17\,000 AU. This suggests that the initial grouping of the ALMA cores is not totally erased over the core life-time, nor the protostar life-time \citep[$\sim$0.5 Myr; e.g.,][]{dun14}. However, the different grouping of stars with disks shows that it is erased over the life-time of stars with disks \citep[$\sim$2 Myr; e.g.,][]{eva09, dun14}. This provides support for the dynamical slingshot model of \citet[][]{stu15prep}, in which the gravitational potential well of the surrounding gas reservoir in the ISF holds the protostars close to their maternal filament section. However, the (proto-)stars decouple from the gas reservoir due to the slingshot, at which point they can loose their initial grouping properties. We note a general caveat that our data probe a narrow field centred on the densest part of the filament. The disk population extends to a significantly larger volume than this field, occupying also a larger volume than dense cores and protostars do. We are insensitive to the grouping signatures the disk population may show outside this field. 

\subsection{Applicability of gravitational fragmentation models}
\label{subsec:jeans}	

We now discuss the observed fragmentation of the ISF in the context of analytic, idealised gravitational fragmentation models. First, shortly recall the physical properties of the ISF. The line-mass of the ISF is $385 \times (D/\mathrm{pc})^{3/8}$ M$_\odot$ pc$^{-1}$, where $D$ is the width of the region that is included in the mass calculation. The radial density structure of the ISF is proportional to $r^{-13/8}$ \citep{stu15prep}. The CO line emission from the ISF shows linewidths around 3-4 km s$^{-1}$ and the C$^{18}$O emission around 1 km s$^{-1}$ when averaged over a three-arcminute beam \citep{nis15}; this indicates strong non-thermal motions given the typical temperatures of $\sim$20 K \citep[e.g.,][]{stu15}. The Zeeman-splitting measurements suggest that the filament is wrapped in a helical magnetic field \citep[][see also discussion in \citealt{stu15prep}]{hei97}. In this paper, we have shown that the filament is fragmented hierarchically into groups of dense cores and that the fragmentation within these groups increases down to 1\,000 AU scales.

Consider now the gravitational fragmentation models for self-gravitating, non-magnetized, infinitely long, equilibrium cylinders in the regime of small perturbations \citep[e.g.,][]{ost64, nag87, inu92, fie00b, fis12}. These models have established that there is a range of unstable wavelengths that lead to growing perturbations. Within this range the perturbation growth rate has a unique maximum that defines the fastest-growing wavelength. Similarly in the spherical case, gravitational fragmentation is expected at the Jeans' length scale \citep{jea29}, provided that sufficiently strong initial density fluctuations are present \citep[e.g.,][]{lar85}. The fundamental prediction of both cylindrical and spherical models is a well-defined fragmentation scale: the fastest growing perturbation wavelength for cylinders and the Jeans' length for spheres. In both cases, this scale is coupled to the gas density.

We immediately notice problems in the applicability of the idealised cylindrical fragmentation models. Most pressingly, the ISF is gravitationally super-critical by a factor of roughly 20, and thus, not near-critical. The applicability of the near-equilibrium cylindrical models to highly super-critical filaments (such as the ISF) has not been established. It has been argued on qualitative grounds, and commonly assumed in literature, that non-thermal motions can provide microturbulent pressure support to super-critical filaments, effectively increasing their critical line-mass \citep[e.g.,][]{fie00a, jac10, fis12, hei13, her14, beu15a}. However, we need to recognise that the nature of non-thermal motions within highly super-critical filaments (including the ISF) has not been established. Observations of nearby, slightly super-critical filaments whose velocity structure has been scrutinised in high detail do not support the picture of micro-turbulent motions within filaments \citep{hac13}. Recent observations suggest analogous, systematic motions also within highly super-critical filaments \citep[e.g.][]{hen14, beu15a}. Furthermore, the radial density distribution of the ISF is $\rho \propto r^{-1.6}$ \citep{stu15prep}, flatter than that of the idealised hydrostatic solution ($\rho \propto r^{?4}$; \citealt{ost64}). Theoretical works indicate that near-critical filaments wrapped by helical magnetic fields \citep{fie00a}, or collapsing filaments with a non-isothermal equation of state \citep{kaw98}, can have flat radial profiles close to $\rho \propto r^{-2}$. However, the applicability of these models to highly super-critical filaments is not clear. Finally, the fragmentation of the ISF is not in agreement with the fragmentation models of near-critical filaments. The ISF does not show a single fragmentation scale, but a more complicated, multi-scale picture: an increase of short separations below 17\,000 AU, especially below 6\,000 AU, and periodic grouping at $\sim$50\,000 AU.

Given the above, the ability of the currently existing analytic, idealised models to address the fragmentation of highly super-critical filaments such as the ISF is questionable at best. This calls for theoretical works aiming at explaining multi-scale, hierarchical fragmentation. In the following, we compare the predictions of the gravitational fragmentation models with the observed fragmentation pattern of the ISF with this short-coming in mind.

\subsection{Comparison against gravitational fragmentation models}
\label{subsec:jeans2}	

In the absence of a theory well-suited for highly super-critical filaments we calculate the predictions of the spherical and cylindrical gravitational fragmentation models for the fragmentation scale of the ISF.  Spherical gravitational fragmentation predicts preferential fragmentation at the Jeans' length-scale, $ l_\mathrm{J} = c_\mathrm{s} (G/\overline{\rho})^{1/2}$, where $c_\mathrm{s}$ is the isothermal sound speed. To evaluate this, we adopt the temperature of 20 K as before. \citet[][]{stu15prep} have evaluated the (cylindric) density profile of the ISF to be 17 M$_\odot$ pc$^{-3} (r / \mathrm{pc})^{-13/8}$ down to their resolution limit of 0.04 pc. If we use the density at 0.04 pc as the proxy of the density relevant for Jeans' fragmentation, the density of $\sim10^5$ cm$^{-3}$ and the fragmentation scale of $\sim$10\,000 AU follows. We also make another estimate of the relevant density by estimating the mean densities of the six main structures in the \emph{Herschel}-derived column density map of the area (see Fig. \ref{fig:2point-sim}). We do this by summing up all column densities in the area of the structures and assuming the structures are spherical. This resulted in the mean density of $2\times 10^5$ cm$^{-3}$, and from therein, in the fragmentation scale of $\sim$7\,000 AU. These fragmentation scales are similar to the size scales over which the cores show an excess of separations, i.e., <17\,000 AU. The scales are smaller by factors of six and eight compared to the 55\,000 AU separation between the core groups identified from the ALMA data (or the separation between the dust structures identified from the \emph{Herschel} data, cf., Fig. \ref{fig:2point-sim}). Thus, from this point-of-view, the fragmentation at $\sim$10\,000 AU scales is in a rough agreement with the prediction of Jeans' fragmentation, while the fragmentation at larger scales seems less likely to be.

What is the prediction of the cylindrical fragmentation models? In the gravitational fragmentation of a near-equilibrium filament (i.e., cylinder), the wavelength of the fastest growing unstable mode depends on the $FWHM$ of the filament, or equivalently, on the central density, scale-radius, or line-mass (these are all inter-connected in the equilibrium solution). The fragmentation scale is then $\sim$3 times the $FWHM$ of the filament \citep[][see also \citealt{nag87, inu92, fie00b}]{fis12}. We do not know well the $FWHM$ of the ISF, but we can use the dependency of the $FWHM$ on central density to obtain an estimate. For a near-equilibrium filament, $FWHM \approx 4 \times c_\mathrm{s} (4 \pi G \rho_\mathrm{c})^{-0.5}$ \citep{fis12}. If we again use the density at $r=0.04$ pc as the relevant density, the $FWHM$$\approx$0.06 pc follows, yielding the fragmentation scale of $\sim$40\,000 AU. If we use the density derived from \emph{Herschel} data (as above), the scale is $\sim$30\,000 AU. These are within a factor of 2 of the periodic separation between the ALMA-detected core groups, or between the \emph{Herschel}-detected dust structures ($\sim$55\,000 AU). Thus, one can speculate that filamentary gravitational fragmentation may be one process affecting the fragmentation at these scales, even if the models do not capture all relevant physical processes affecting fragmentation. Several other works have also found fragmentation in highly-supercritical filaments at these scales \citep[e.g.,][]{bus13, kai13, tak13, wan14, beu15b, tei15, hen16}. One should note that, e.g., the presence of magnetic fields can significantly alter the prediction for the fragmentation scale \citep[e.g.,][]{fie00b}. We also have assumed an inclination angle of zero throughout this paper. 


In summary, the observational picture of fragmentation within ISF is not predicted by any single existing fragmentation model alone, even if the models capture some characteristics of the observed fragmentation. To make progress, a multi-scale model specific for highly super-critical filaments is needed. Since the models for near-critical filaments are partially in agreement with observations, they may provide a reasonable basis for such models. We can speculate about such framework in the light of the fragmentation time-scales predicted by the existing models. A high-aspect ratio filament first fragments following the prediction of the cylindrical gravitational fragmentation model (in the ISF, this would correspond to the fragmentation seen at 50\,000 AU scale). The characteristic time-scale for this is $\tau_\mathrm{m} \approx 3 \times \sqrt{1/(4\pi \rho_\mathrm{c} G)} \approx 0.2$ Myr. By definition, the fragments have a higher mean density than the initial filament (assuming no mass is lost to stars at this point). This leads to a shorter fragmentation time-scale within the fragment compared to the initial fragmentation time, by a factor of $\rho_\mathrm{c}^{-1/2}$ \cite[e.g.,][]{fie00b, fis12}. If the global collapse time-scale of the fragments is their free-fall time (about 0.1 Myr at the density of $10^5$ cm$^{-3}$), this leads to a competition between the global collapse of the fragment and further fragmentation inside it \citep[e.g.,][]{pon11, pon12, cla15}. Clearly, the latter must ensue faster, because we observe fragmentation down to $\sim$1\,000 AU scales. However, it is not obvious how to arrive to this conclusion, because both the free-fall time and the fragmentation time depend on density with the same power ($\propto \rho^{-0.5}$). One possible route is provided by magnetic fields that may provide support against a global collapse while not strongly affecting the fragmentation time-scale \citep[e.g.,][see also \citealt{sei15} for a numerical study showing an analogous effect in filaments that are thermally near-critical]{fie00a, fie00b}. Another possibility could be large-enough, pre-existing density fluctuations that can grow and collapse faster than the global, longitudinal collapse ensues \citep{lar85, pon11}. Supporting this possibility, filaments with low star formation activity have been observed to contain significant density fluctuations \citep[e.g.,][]{roy15, kai16}. Developing this cartoon framework into a coherent theory is beyond this paper, but it would be a crucial topic for future works that aim at understand the fragmentation of highly super-critical filaments. 

 \section{Conclusions}

We present a fragmentation study of a 1.6 pc long section of the ISF in the Orion A molecular cloud, specifically covering the OMC-2 and a part of the OMC-3 region. We employ ALMA 3 mm continuum emission data that reach the spatial resolution of $\sim$3", or 1\,200 AU. We also study the relationship between the distributions of the ALMA-detected dense cores, protostars, and stars with disks using the census of the young stellar population built by the \emph{Spitzer} Orion Survey and HOPS. Our conclusions are as follows.

\begin{enumerate}

\item The ALMA data reveal numerous substructures from the ISF, including compact and extended structures and filaments. Most of the single-peaked structures detected previously in coarser resolution resolve into substructures. Especially, we detect a prominent filament in the  FIR 3-5 region connecting a series of dense cores and protostars.

\item We identify 43 dense cores from the ALMA data, 26 (60\%) of which coincide with protostars and 17 (40\%) of which are starless. The nearest neighbour separation distribution of the ALMA cores increases down to our resolution limit, indicating that there is no preferential nearest-neighbour separation. The nearest-neighbour separation distributions of the starless cores and protostellar cores are different; the starless cores show a higher fraction of short nearest-neighbour separations than the protostellar cores.

\item The two-point correlation analysis of the dense core separations shows that the ALMA cores are significantly grouped at separations smaller than $\sim$17\,000 AU and strongly grouped at separations smaller than $\sim$6\,000 AU. This is different for starless and protostellar cores: the starless cores show strong grouping, while the protostellar cores show only marginally significant grouping, however, the grouping strength of the cores (especially starless cores) may be affected by caveats in the core identification. The increase of grouping below 6\,000 AU may be related to the regime where interactions between the cores become abundant. The spatial distribution of the cores also indicates a quasi-periodic grouping of the cores into groups $\sim$30\,000 AU in size, separated by $\sim$50\,000 AU. These groups corresponds to the gas morphology traced by the \emph{Herschel} column density map of the region \citep{stu15}. 

\item The two-point correlation analysis shows that the HOPS protostars in the entire northern ISF are significantly grouped at separations shorter than $\sim$17\,000 AU. Thus, they show partially similar grouping as the dense cores; however, the protostars do not show strong grouping below $\sim$6\,000 AU like the dense cores do. The distribution of stars with disks is in agreement with random distribution within the ALMA-covered region, and it is thus significantly different from that of the ALMA cores and protostars. These results suggest that the grouping of dense cores, resulting from the fragmentation process, is not totally erased during the protostar life-time, but it is erased during the longer life-times of stars with disks. This is in agreement with the picture of \citep{stu15prep} in which the protostars are ejected from the filament by the slingshot mechanism.  

\item The hierarchical, scale-dependent fragmentation we observe in the ISF is not self-consistently predicted by any existing gravitational fragmentation models. We use the density profile along the filament derived by \citet{stu15prep} and show that the predictions for the Jeans' fragmentation scale is $\sim$10\,000 AU and for the filamentary gravitational fragmentation scale is $\sim$40\,000 AU. These scales are similar to the scales at which the cores show grouping, suggesting that gravitational fragmentation is an important process, even though our understanding of how exactly it proceeds is incomplete.

\end{enumerate}

Our results provide the most sensitive view of the $1\,000$ AU scale fragmentation within a massive filament to date. Especially, our detection of increasingly abundant separations below 6\,000 AU opens the question of the abundance of bound groups at these separations. Further observations targeting the dynamics of the cores and scales between 500-1\,000 AU will be able to quantify this and address the question of stellar multiplicity resulting from core fragmentation. Similarly, our work highlights the need for theoretical work addressing the structure of highly super-critical filaments and the multi-scale nature of their fragmentation; our observations provide concrete constraints for testing such models.


\begin{acknowledgements}
The authors are grateful to Ralph Pudritz and Jaime Pi\~neda for fruitful discussions. 
The work of J.K. and A.S. was partially supported by the Deutsche Forschungsgemeinschaft priority program 1573 ("Physics of the Interstellar Medium"). 
This project has received funding from the European Union's Horizon 2020 research and innovation programme under grant agreement No 639459 (PROMISE). 
This paper makes use of the following ALMA data: ADS/JAO.ALMA\#2013.1.01114.S. ALMA is a partnership of ESO (representing its member states), NSF (USA) and NINS (Japan), together with NRC (Canada), NSC and ASIAA (Taiwan), and KASI (Republic of Korea), in cooperation with the Republic of Chile. The Joint ALMA Observatory is operated by ESO, AUI/NRAO and NAOJ.
\end{acknowledgements}



\appendix


\section{Effect of the background filter width on the core identification}
\label{app:bgtest}

\begin{figure}
   \centering
\includegraphics[width=\columnwidth]{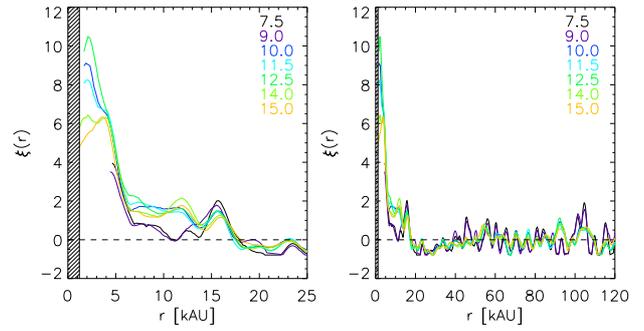}
      \caption{Two-point correlation functions of the dense cores populations identified with different background filter widths of the \textsf{gaussclumps} algorithm. The left panel shows the function between 0-25 kAU and the right panel between 0-120 kAU. The coloured numbers show the background filter width, in arcseconds, used in the core identification. 
              }
         \label{fig:2point-bg-test}
   \end{figure}

We tested the effect of the background filter width used in the dense core identification on the resulting two-point correlation functions. We repeated the core identification with the filter widths between 7.5$\arcsec$-15$\arcsec$. The resulting two-point correlation functions are shown in Fig. \ref{fig:2point-bg-test}. The functions are drawn only above the minimum separation of the dense cores. The test shows that the choice of the background filter does not affect the main conclusions derived from the two-point correlation functions. This comes from the fact that the majority of the cores are identified similarly regardless of the choice of the background filter width. When a small filter is chosen, only the strongest (most significant) cores are detected. When the filter width approaches 15$\arcsec$, structures identified as separate dense cores are occasionally identified as one, larger core. The test shows that optimum filter width for detecting cores in 3$\arcsec$ resolution is between $\sim$7.5$\arcsec$-12.5$\arcsec$. Our final choice, 10$\arcsec$, is within this range.

\section{Two-point correlation analysis at large scales}
\label{app:2point}


We estimated the significance of the quasi-periodic core grouping at 50\, 000 scales with a simple simulation. The simulation setup was guided by the surface number density of the ALMA-detected cores (Fig. \ref{fig:2point-sim}). We construct a simplistic model in which  groups of cores are located along a line with a constant separation. The simulation area matches the dimensions of the ALMA field, and we place six groups in the field. The groups have the size of 42\,000 AU and their separation is 56\,000 AU. The placement of cores within the groups is random. The bottom right panel of Fig. \ref{fig:2point-sim} shows a typical two-point correlation function resulting for 43 simulated cores. The simulation illustrates that the basic signatures of this configuration, i.e., the deficit of cores at $D$ (42\,000 AU) and the excess at $\lambda$ (56\,000 AU), are generally expected to be detectable with a sample of 43 cores, albeit with a low significance.

  \begin{figure}
   \centering
\includegraphics[bb = 300 5 650 330, clip = true, width=0.75\columnwidth]{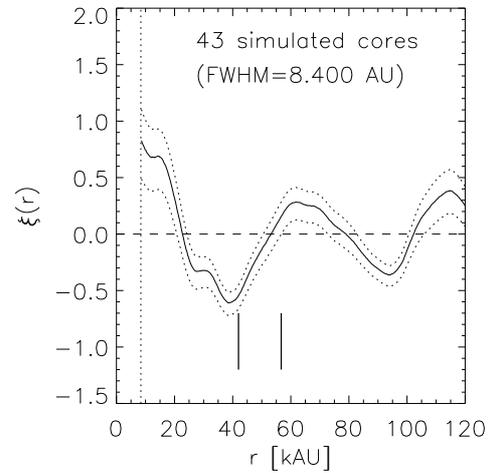}
      \caption{Two-point correlation function of a simulation in which 43 cores are randomly placed in six groups along a line. The diameter of the groups is 42\,000 AU and their separation 56\,000 AU. These distances are indicated with vertical lines. The function was derived with a smoothing function that has $FWHM=20\arcsec = 8.400$ AU.
              }
         \label{fig:2point-sim-appendix}
   \end{figure}

\section{On the robustness of the two-point correlation analysis}
\label{app:robustness}

\begin{figure}
   \centering
\includegraphics[width=\columnwidth]{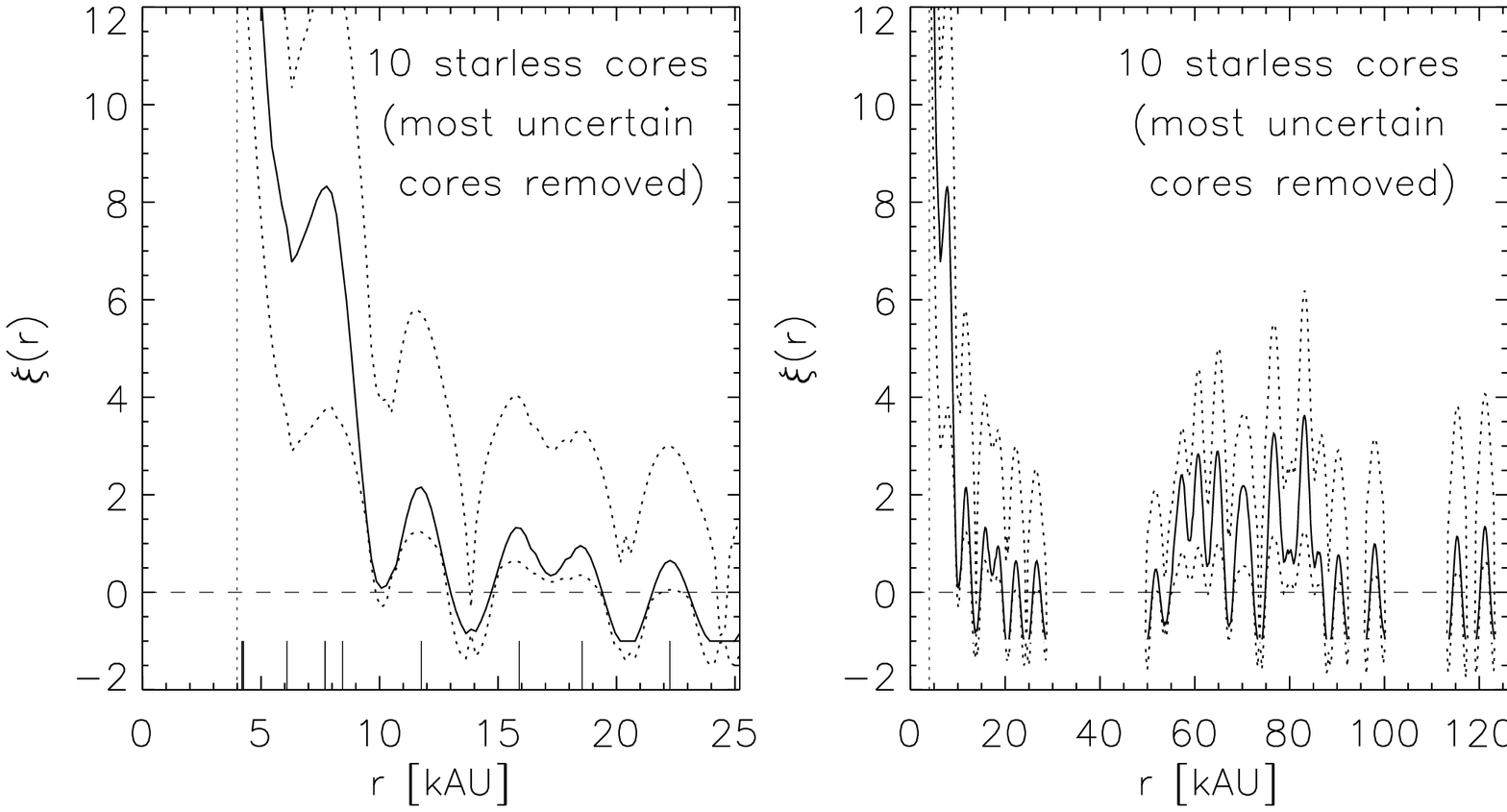}
\includegraphics[width=\columnwidth]{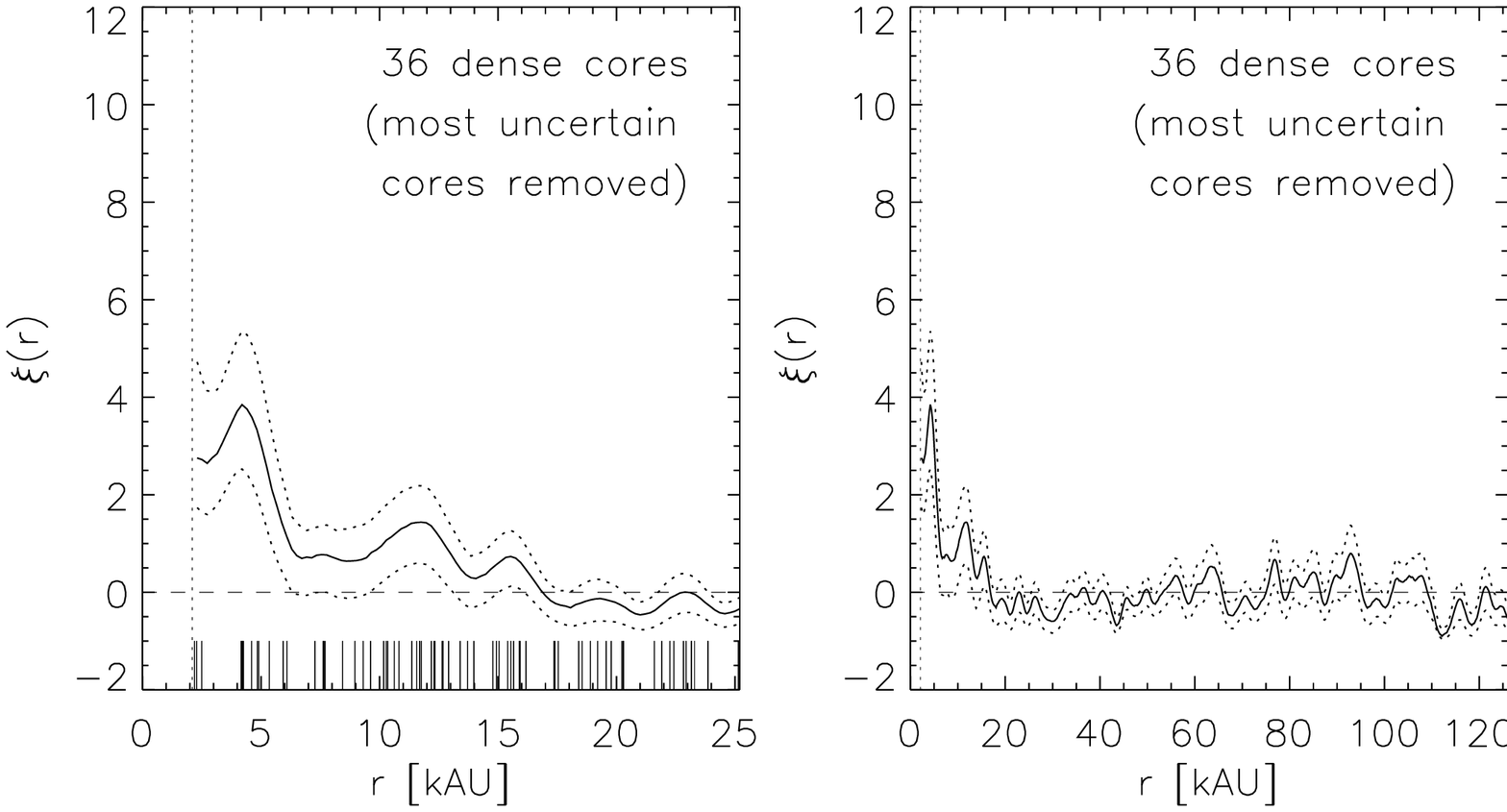}
      \caption{Two-point correlation function of the separations of all ALMA dense cores and starless cores, excluding the seven most uncertain starless cores. In the top right panel, there are not enough separations to compute the two-point correlation function at all scales.
              }
         \label{fig:2point-app2}
   \end{figure}

The continuum data we exploit in this data does not alone allow  disentangling cores and core-like objects (e.g., walls of outflow cavities, structures overlapping along the line-of-sight) from each others. Consequently, especially the weakest structures classified as starless cores may be misclassifications. To address the effect of this possibility on the two-point correlation analysis, we derived the two-point correlation functions of starless cores and all dense cores excluding seven weakest starless cores (\#5, 15, 18, 19, 23, 26, 35). The resulting two-point correlation functions are shown in Fig. \ref{fig:2point-app2}. The two-point correlation function for starless cores only includes ten cores and is very uncertain. However, it do shows a significant excess of short separations ($\lesssim$ 10\,000 AU). The two-point correlation function for all dense cores shows significant excess of separations below $\sim$15\,000 AU. The function increases down to our resolution limit, albeit less strongly as when all identified dense cores are included (Fig. \ref{fig:2point-cores}). We conclude that the results based on two-point correlation analysis are not strongly affected by possible misclassifications of low-intensity starless cores.

\end{document}